\documentclass[twoside]{ilcws08}
\usepackage[latin1]{inputenc}
\usepackage[dvips]{graphicx,epsfig,color}
\usepackage{wrapfig,rotating}
\usepackage{amssymb,amsmath,array}

\renewcommand{\thesubsection}{\thesection.\alph{subsection}}

\newcommand{\be}{\begin{equation}}
\newcommand{\ee}{\end{equation}}
\newcommand{\bea}{\begin{eqnarray}}
\newcommand{\eea}{\end{eqnarray}}
\newcommand{\gsim}{\hbox{ \raise3pt\hbox to 0pt{$>$}\raise-3pt\hbox{$\sim$} }}
\newcommand{\lsim}{\hbox{ \raise3pt\hbox to 0pt{$<$}\raise-3pt\hbox{$\sim$} }}
\newcommand{\mathbold}[1]{\mbox{\boldmath $\bf#1$}}

\def\to{\rightarrow}
\newcommand{\figdir}{figs}

\begin{document}
\title{
Feasibility study of the forward-backward\\
asymmetry of the $\mathbold{\it{e^+e^- \rightarrow t\bar{t}}}$
process\\
in all-hadronic decay modes
at $\mathbold{\it{\sqrt{s} = {\rm{500}}}}$ GeV\\
with the ILD detector
}

\author{
Katsumasa Ikematsu\footnote{Corresponding author.
\ e-mail address: \texttt{Katsumasa.Ikematsu@kek.jp}},
Akiya Miyamoto, and Keisuke Fujii
\vspace{.3cm}\\
High Energy Accelerator Research Organization (KEK),\\
Tsukuba, 305-0801, Japan
}

\maketitle

\begin{abstract}
We have studied the measurement accuracy
of the forward-backward asymmetry
of the $e^+e^- \rightarrow t\bar{t}$ process
in the 6-jet mode at $\sqrt{s} = 500$ GeV
with the ILD detector.
In the analysis the vertex charges of $b$-jets were used
to distinguish $t$ from $\bar{t}$ in each event.
The distribution of the cosine of the reconstructed polar angle
of so identified $t$ or $\bar{t}$
showed a clear forward-backward asymmetry.
After the correction for charge misidentification
the forward-backward asymmetry was determined to be
$A^t_{FB} = 0.334 \pm 0.0079$
for 500 fb$^{-1}$
with the beam polarization combination of
$P(e^+, e^-) = (+30\%, -80\%)$,
demonstrating a very good statistical accuracy ($\sim 2\%$) even
in the 6-jet mode.
\end{abstract}

%
%
\section{Introduction}

The forward-backward asymmetry of
the $e^{+}e^{-} \rightarrow t\bar{t}$ process is sensitive to the $t\bar{t}Z$
and $t\bar{t}\gamma$ couplings and serves as a probe for new physics,
which may appear as anomaly in these couplings.

Needless to say we have to distinguish $t$ from $\bar{t}$
in each event in order to measure the forward-backward asymmetry, $A^t_{FB}$.
In the lepton+4-jet mode, it is straightforward
because the lepton charge tells the charge of
the leptonically decayed $W$,
and hence identifies its parent to be either $t$ or $\bar{t}$.
In the 6-jet mode, however,
we need some other way to separate $t$ and $\bar{t}$,
which is non-trivial.
Nevertheless it is worth investigating the feasibility of
$A^t_{FB}$ measurement in the 6-jet mode,
since it has a major branching fraction of 46\%.
In addition, the kinematical fit works better
in the 6-jet mode than in the lepton+4-jet mode,
where a large energy is taken away by
the neutrino from the leptonically decayed $W$.
The 6-jet mode might, hence, be advantageous
when the influence of beamstrahlung is significant.

In this paper, we use the vertex charges of $b$ jets
to distinguish $t$ from $\bar{t}$ in each event.
The measurement of the vertex charges requires
a high performance detector system as well as
a sophisticated reconstruction algorithm.
The forward-backward asymmetry in the 6-jet mode is
therefore a very good measure for the overall performance
of a detector system,
hence being included as one of the ILC LoI benchmarks
~\cite{Ref:ILCBenchmark}.
The benchmark conditions are
the center of mass energy of $\sqrt{s}=500$ GeV,
an integrated luminosity of $500 \, {\rm fb}^{-1}$,
and a beam polarization combination of
$P(e^+, e^-) = (+30\%, -80\%)$.

The ILD~\cite{Ref:ILDLOI} is equipped with
an unprecedentedly excellent vertex detector,
which allows efficient $b$-jet charge identification
with the vertex charge.
The vertex charges of $b$ jets were reconstructed by
the LCFIVertex algorithm~\cite{Ref:LCFIVertex}.

This paper is organized as follows.
After we briefly describe our analysis framework
in section~\ref{Sec:Framework},
we move on to the vertex charge reconstruction
and show how well we can identify each jet as $b$ or $\bar{b}$
in section~\ref{Sec:BottomID}.
We then apply this to $t/\bar{t}$ identification for
the determination of the production angle distribution
($dN / d\cos \theta_{t}$) in section~\ref{Sec:TopAngle}.
After the correction for charge misidentification
we derive the $A^t_{FB}$ and discuss the result
in comparison with the Monte Carlo truth in section~\ref{Sec:TopAFB}.
Section~\ref{Sec:Summary} summarizes our analysis
and concludes this paper.

%
%
\section{Analysis framework}
\label{Sec:Framework}

In general, Monte Carlo (MC) simulation consists of the following steps:
event generation, detector simulation, event reconstruction,
and data analysis.

All of the MC event samples,
both the signal and the backgrounds, used in this study
were produced in the StdHep~\cite{Ref:StdHep}
format by a SLAC team~\cite{Ref:SLACLoISamples}
as common inputs to LoI studies,
using WHIZARD 1.4~\cite{Ref:Whizard}
for generating parton 4-momenta and PYTHIA 6.2~\cite{Ref:Pythia}
for parton-showering and hadronization.
The beam energy spread and the beamstrahlung were
properly taken into account using
the spectrum generated with Guinea-Pig~\cite{Ref:GuineaPig}
for the default ILC design parameters at $\sqrt{s} = 500 \, {\rm GeV}$.

The final-state particles output in the StdHep format
from the event generation step were passed to
a Geant4-based detector simulator called Mokka~\cite{Ref:Mokka}
and swum through the ILD detector to create exact hits
in trackers and calorimeters.

These exact hits were smeared or digitized, if necessary,
depending on the detector elements
in the first part of MarlinReco~\cite{Ref:Marlin}~\cite{Ref:MarlinReco}.
The pattern recognition was done for the smeared tracker hits,
separately in the TPC and the silicon detectors
and so found track segments were then linked together
and fed into a Kalman-filter-based track fitter
in the second part of MarlinReco.
From the fitted tracks and the calorimeter hits,
individual particles were reconstructed
as particle flow objects (PFO)
with a sophisticated particle flow algorithm
called PandoraPFA~\cite{Ref:PandoraPFA}
in the third part of MarlinReco.

These PFOs were forced to cluster into 6 jets
for the signal and all background events with
Durham jet clustering algorithm~\cite{Ref:DurhamJet}
in the fourth part of MarlinReco.

The next step is heavy flavour tagging with
LCFIVertex~\cite{Ref:LCFIVertex}.
The LCFIVertex package consists of two parts.
The first part is to search for secondary and tertiary vertices
inside each jet and locate them,
thereby determining the decay length,
the transverse mass, and the momentum at each of these vertices.
In the second part these quantities are translated into
the impact parameter joint probability and the highest two impact
parameter significances,
which are used as inputs into a neural net (NN)
trained with jet samples
having 0 and 1 or more secondary vertices.
Each reconstructed jet is then assigned with the three NN outputs,
corresponding to $b$-, $c$-, and $bc$-tags.

Once a bottom-flavoured jet is identified
we can determine whether it is $b$ or $\bar{b}$
by measuring the vertex charge.
We will discuss this in detail in section~\ref{Sec:BottomID}.

The reconstruction of $e^+e^-\rightarrow t\bar{t}$ events
in 6-jet final states has been studied extensively in the context of
the top quark mass measurement. The top quark mass
in the 6-jet mode is one of the benchmark observables for ILC LoIs
and the ILD study has been reported in reference~\cite{Ref:ILDLOI}.
Similar to the top quark mass measurement, the measurement of the
forward-backward asymmetry requires a correct jet-parton association.
In this paper we employed the same reconstruction algorithm
as those used for the top quark mass measurement.
For the reconstruction, therefore,
we refers the readers to the above reference.

%
%
\section{Vertex charge reconstruction and
         its performance for a single $\mathbold{\it b}$-jet}
\label{Sec:BottomID}

\begin{figure}[htbp]
  \begin{center}
    \includegraphics[height=5cm,clip]{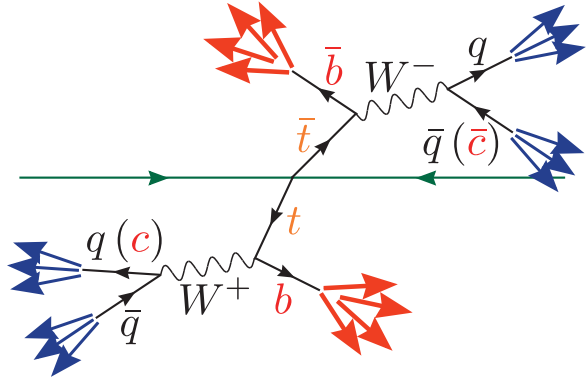}
  \end{center}
  \caption{
        Schematic diagram showing fully hadronic decay channel
        of $e^+e^-\rightarrow t\bar{t}$ process}
  \label{Fig:tt6j-FullHad}
\end{figure}

In principle, we can tell which three jets are from top and
the rest from anti-top in the 6-jet final state
by identifying either a $b/\bar{b}$-jet
or a $c/\bar{c}$-jet from $W^+/W^-$ decay
as shown in Fig.~\ref{Fig:tt6j-FullHad}.
In this analysis, however, we used only $b/\bar{b}$-jets
for the top/anti-top separation.

A bottom quark hadronizes into a $B$-hadron,
which flies over a finite distance thanks to
its large $c\tau$,
making a secondary vertex significantly away
from the primary vertex and hence
identifiable by the vertex detector.
We define the vertex charge as the sum of the charges of
the charged tracks associated with the secondary vertex.
If the charged tracks are reconstructed and associated perfectly,
the vertex charge is equal to the charge of
the primary $B$-hadron, from which
the charge sign of the $b/\bar{b}$ is uniquely determined.

In practice, however, there is no perfect vertexing,
and the resultant distribution of the vertex charges
of charged $B$-hadrons will have a finite width
and hence their charges might sometimes be mis-identified.
By the same token, the charges of neutral $B$-hadrons
might be mis-identified, causing confusion
in bottom charge sign identification.

\begin{figure}[htbp]
  \begin{center}
    \includegraphics[height=6cm,clip]{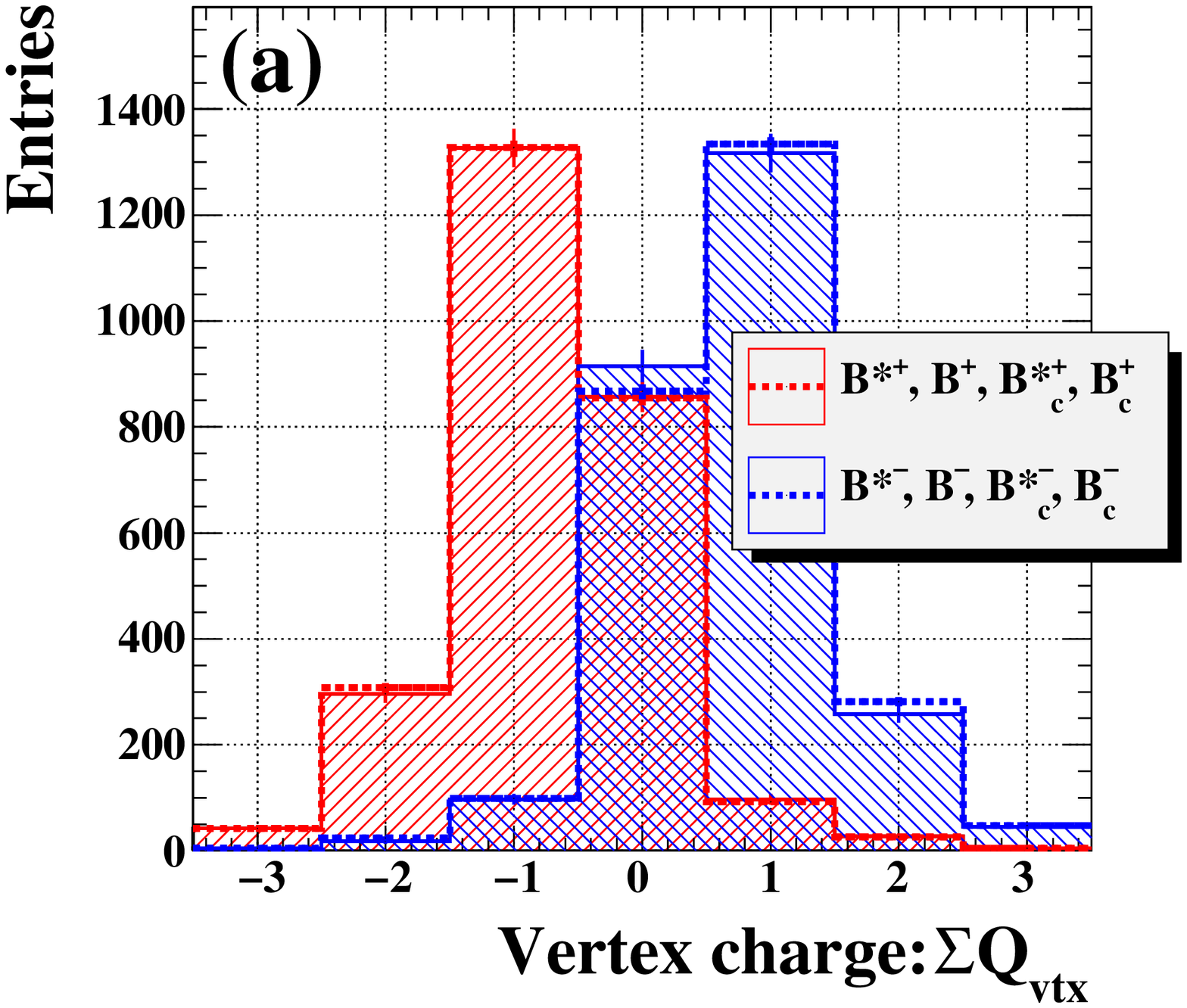}
    \includegraphics[height=6cm,clip]{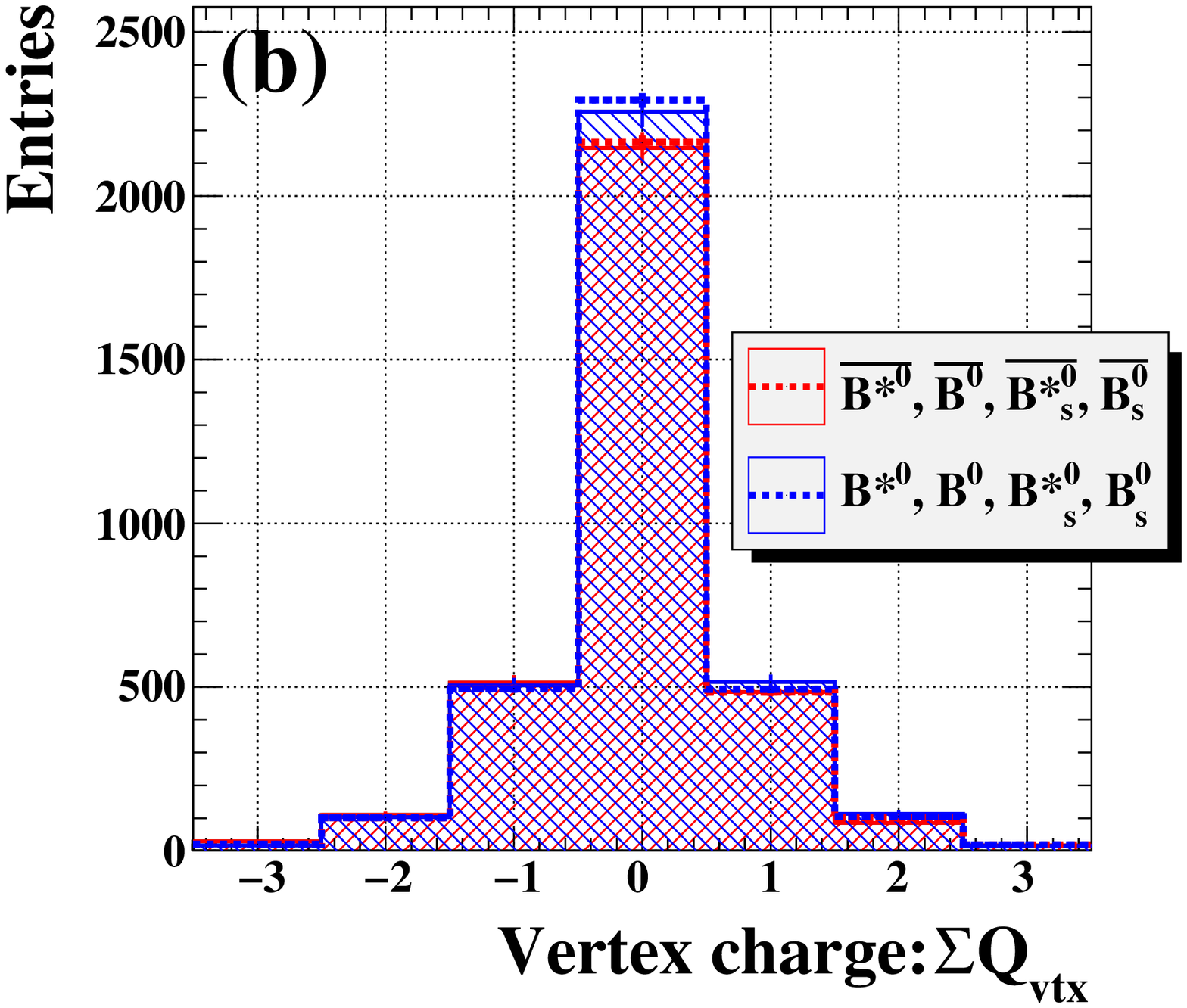}
  \end{center}
  \caption{
        Reconstructed vertex charge distributions
        for (a) charged $B$-mesons and (b) neutral $B$-mesons
        as separated using MC truth information
        (PDG particle ID code).}
  \label{VertexChargeMC}
\end{figure}

Figure~\ref{VertexChargeMC} shows the distribution of the MC level
vertex charge, which is defined by the charge of
a $B$-meson ($B^\pm$ or $B^0$)
involved in the jet tagged as a $b$-jet.

Inspection of the figures tells us that we can separate
$B^+$s from $B^-$s by selecting jets with negative or positive
vertex charges with some contamination from $B^0$s.

\begin{figure}[hbtp]
  \begin{center}
    \includegraphics[height=7cm,clip]{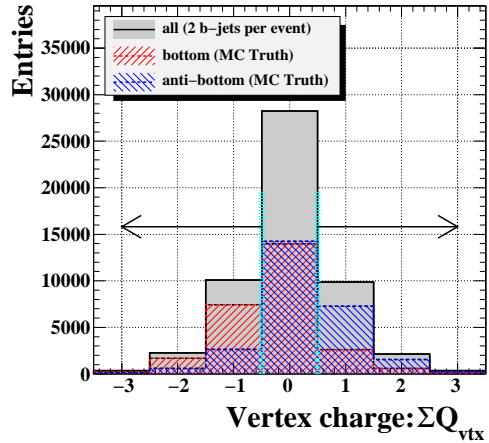}
  \end{center}
  \caption{
        Reconstructed vertex charge distributions
        for $b$-quarks (hatched red) and
        $\bar{b}$-quarks (hatched blue),
        as identified using MC truth information,
        shown together with that of all the reconstructed
        $b$-jets (solid gray).}
  \label{VertexCharge}
\end{figure}

Figure~\ref{VertexCharge} shows the distribution of the
reconstructed vertex charges
of the jets which are $b$-tagged.
In this distribution the $b$-quark ($\bar{b}$-quark)
contribution is shown by hatched blue (red).

The charge sign identification efficiency for a single $b$-jet
using the vertex charge is 28\% with a purity of 75\%.
Notice that only 40\% of the b-jets hadronize into
charged $B$-hadrons and hence maximum efficiency one can hope
for is 40\% in this method.
%
%
\section{$\mathbold{\it t}$ and
$\mathbold{\bar{\it t}}$ identification and
determination of the production angle distribution
($\mathbold{\it dN / d\cos \theta_{t}}$)}
\label{Sec:TopAngle}

\begin{figure}[hbtp]
  \begin{center}
    \includegraphics[height=7cm,clip]{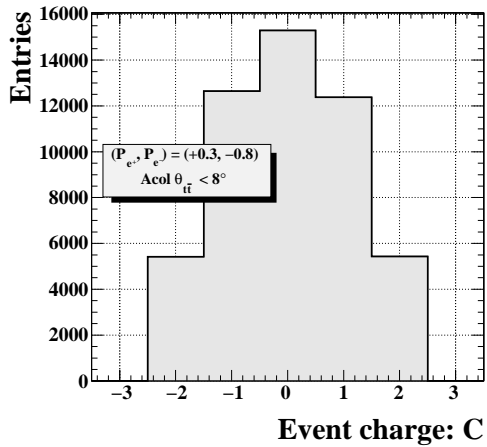}
  \end{center}
  \caption{
        Distribution of event charge: $C = c_1 - c_2$.}
  \label{EventCharge}
\end{figure}

Let's call two reconstructed top systems $t_1$ and $t_2$
and $b$-tagged jets associated to them $b_1$ and $b_2$, respectively,
The identification of $t$ and $\bar{t}$ is performed by using
the vertex charges of $b_1$ and $b_2$ as follows.
We define $c_i$ ($i=1$ and $2$) as the vertex charge of $b_i$ and
$C \equiv c_1 - c_2$ as the event charge.
If $C$ is 0, the event is thrown away as we cannot tell $t$ from $\bar{t}$.
If $C$ is positive, $t_1$ is $t$,
while if $C$ is negative, $t_1$ is $\bar{t}$.
The typical distribution of the event charge $C$
is shown in Fig.~\ref{EventCharge}.

\begin{figure}[hptb]
  \begin{center}
    \includegraphics[height=4.3cm,clip]{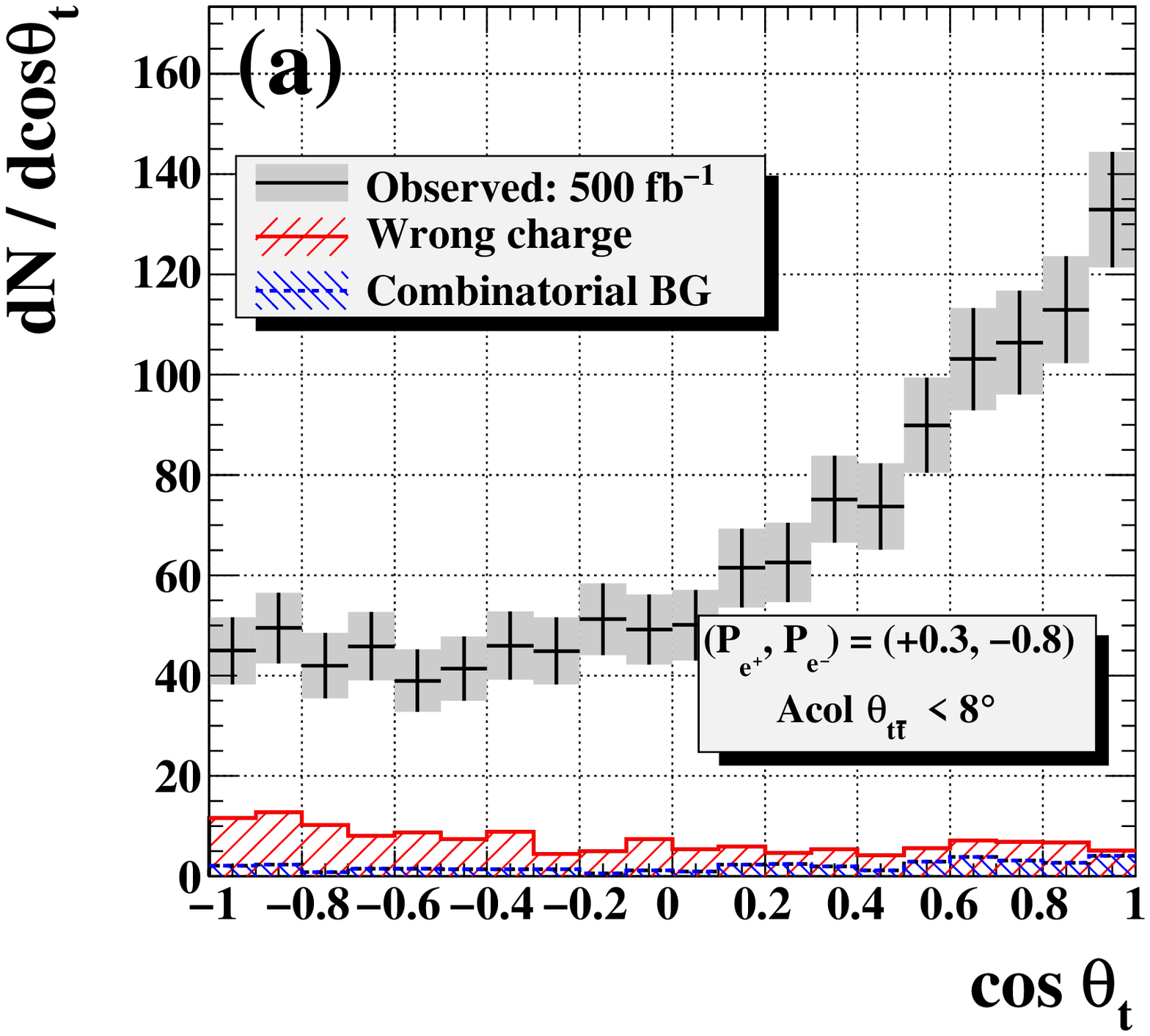}
    \includegraphics[height=4.3cm,clip]{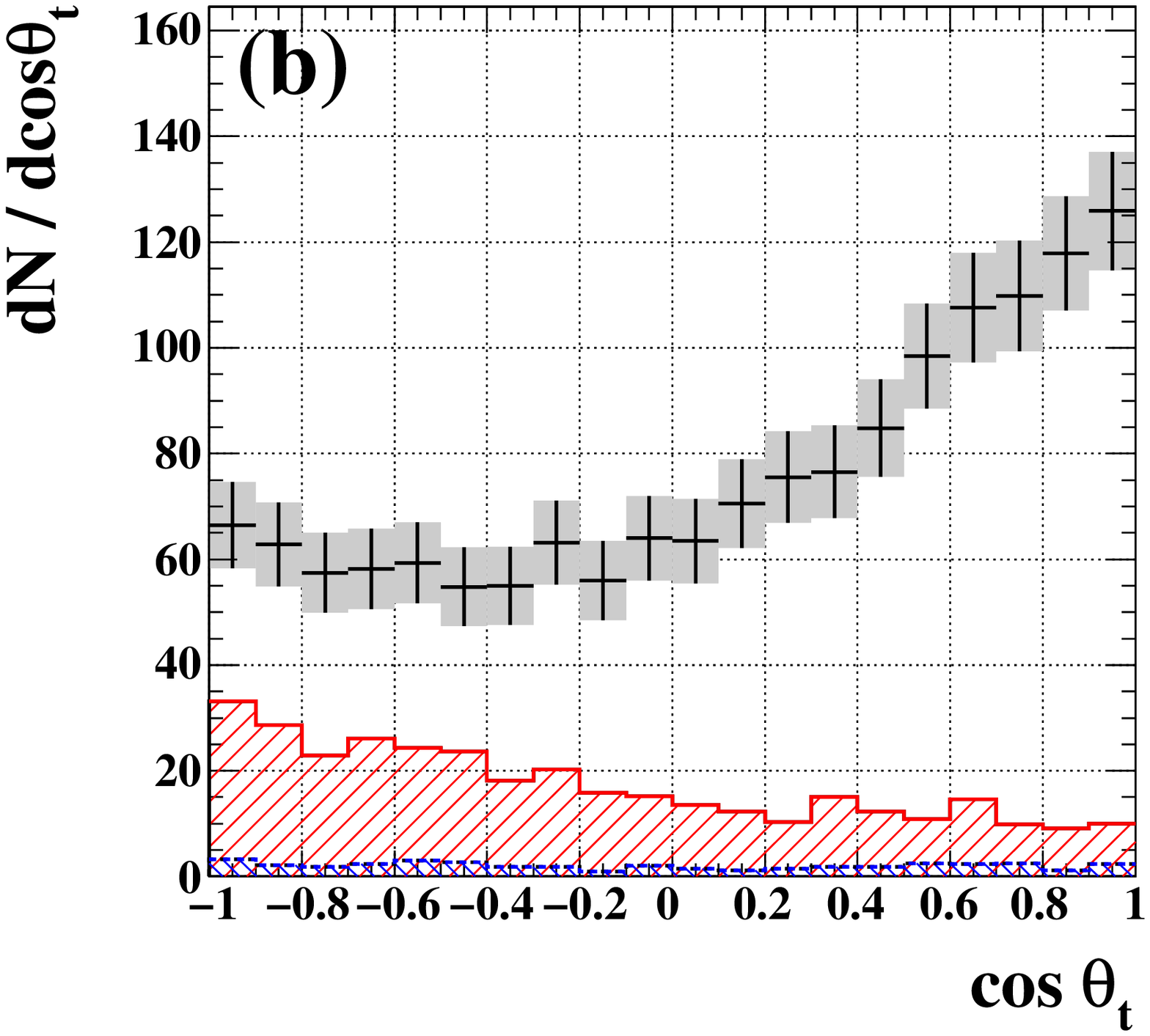}
    \includegraphics[height=4.3cm,clip]{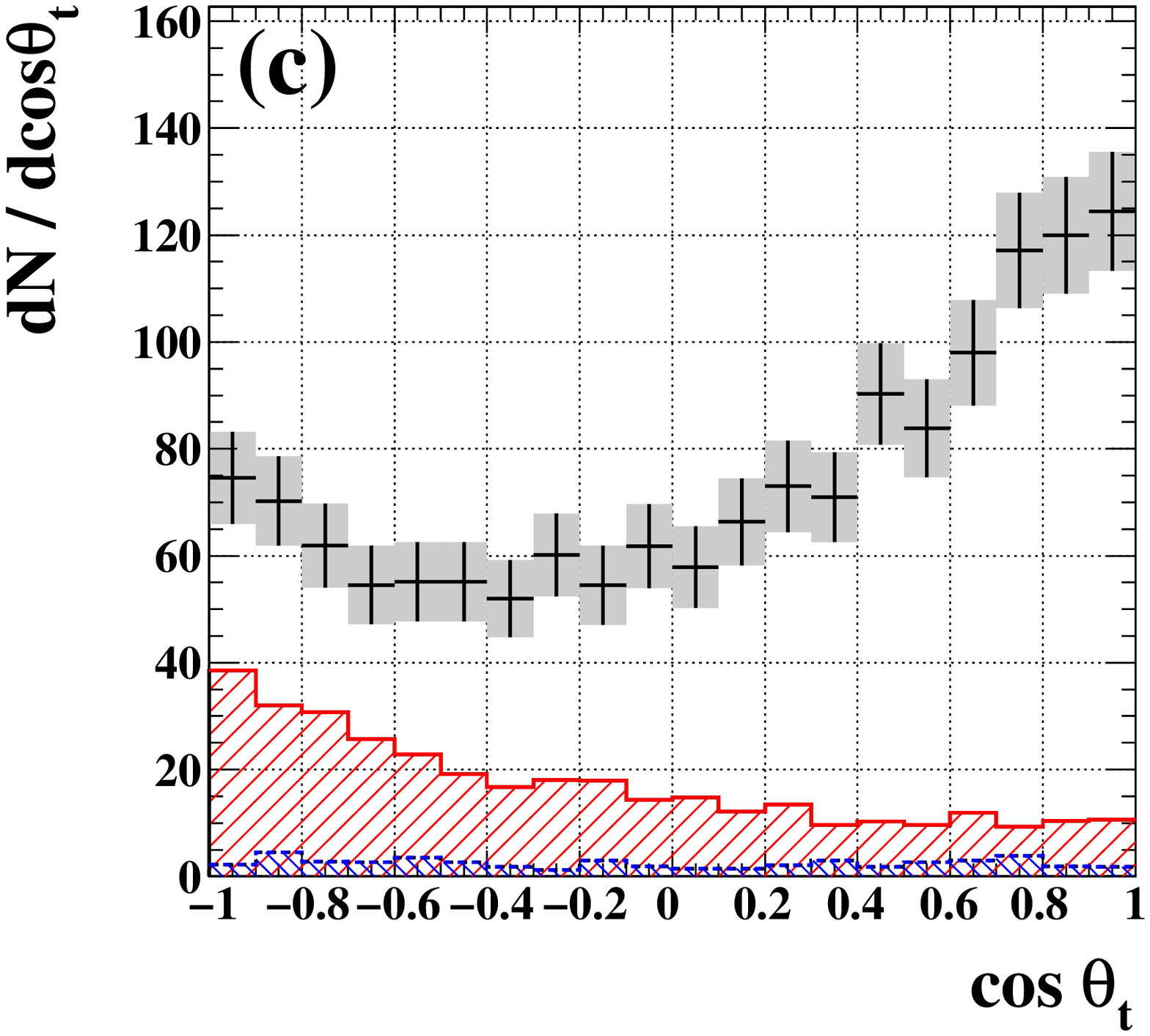}
    \\
    \includegraphics[height=4.3cm,clip]{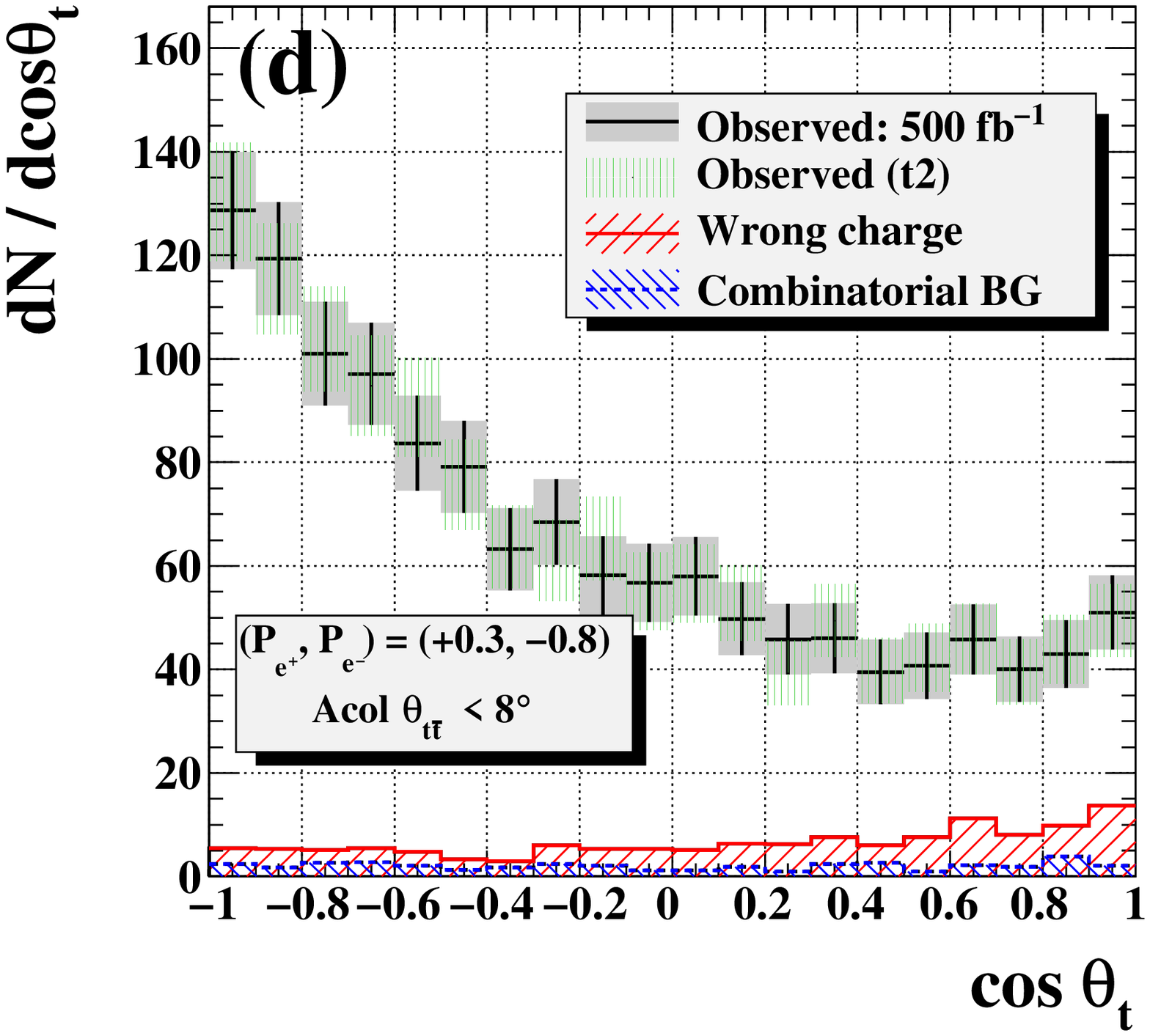}
    \includegraphics[height=4.3cm,clip]{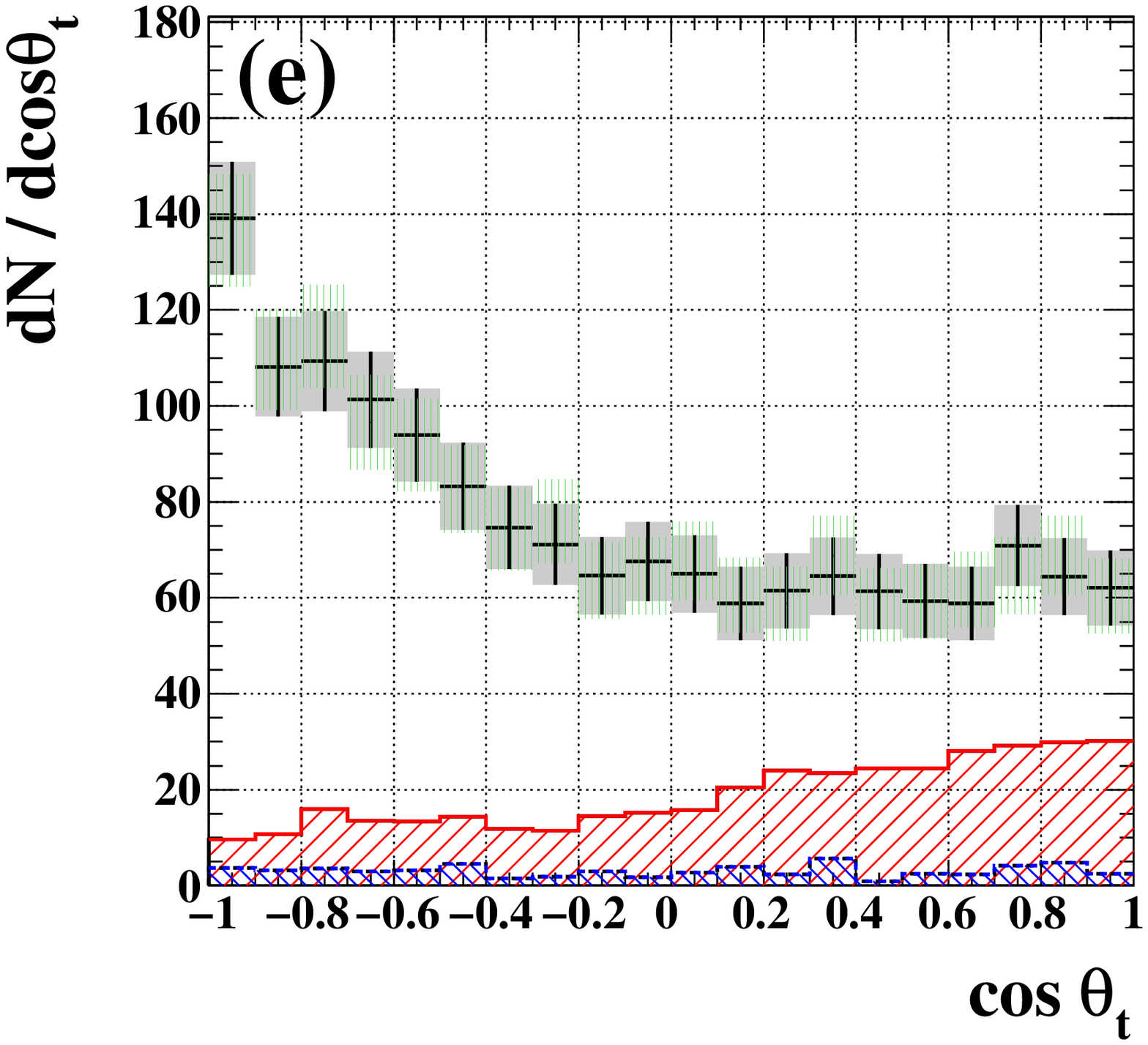}
    \includegraphics[height=4.3cm,clip]{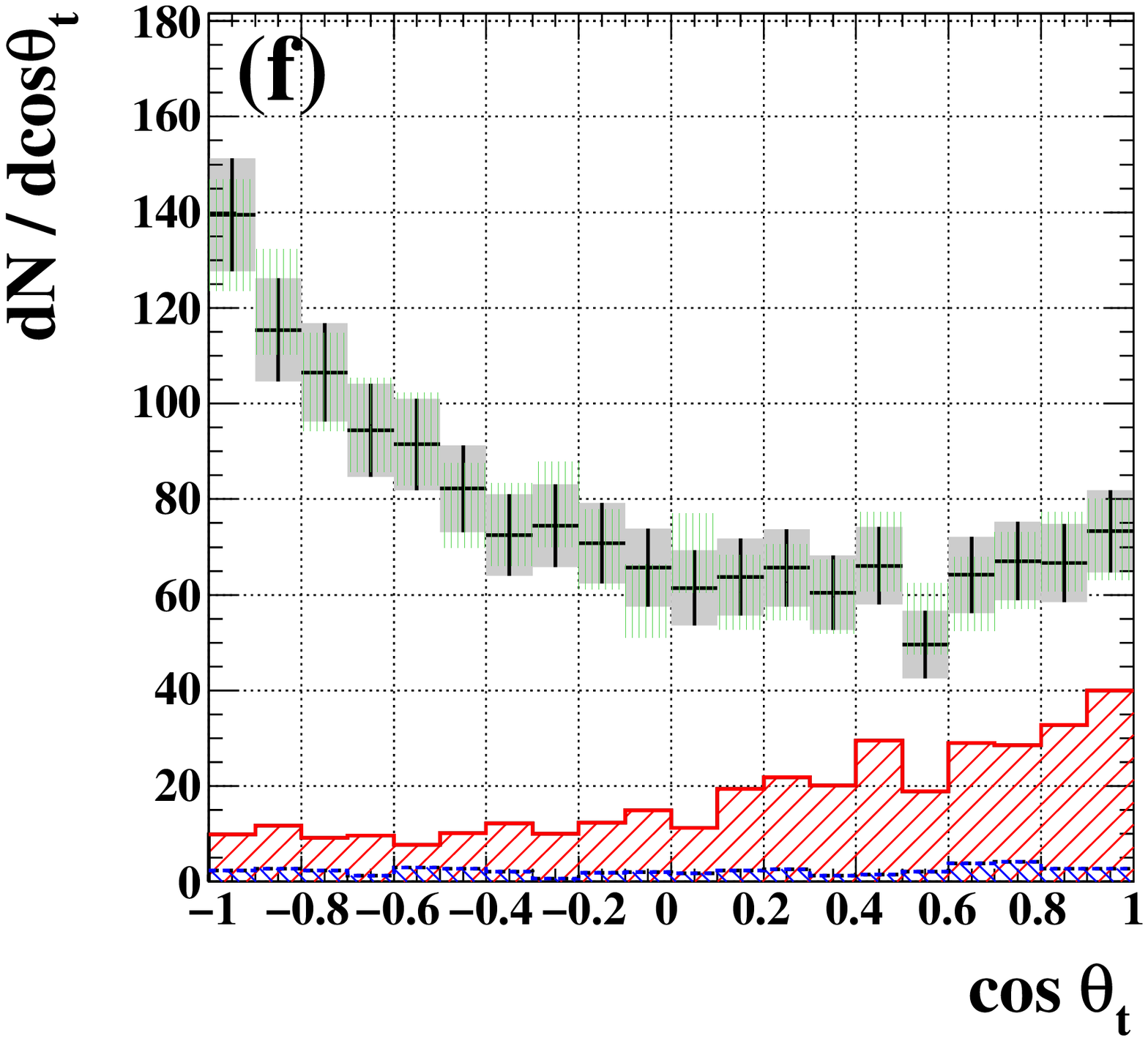}
    \\
    \includegraphics[height=4.3cm,clip]{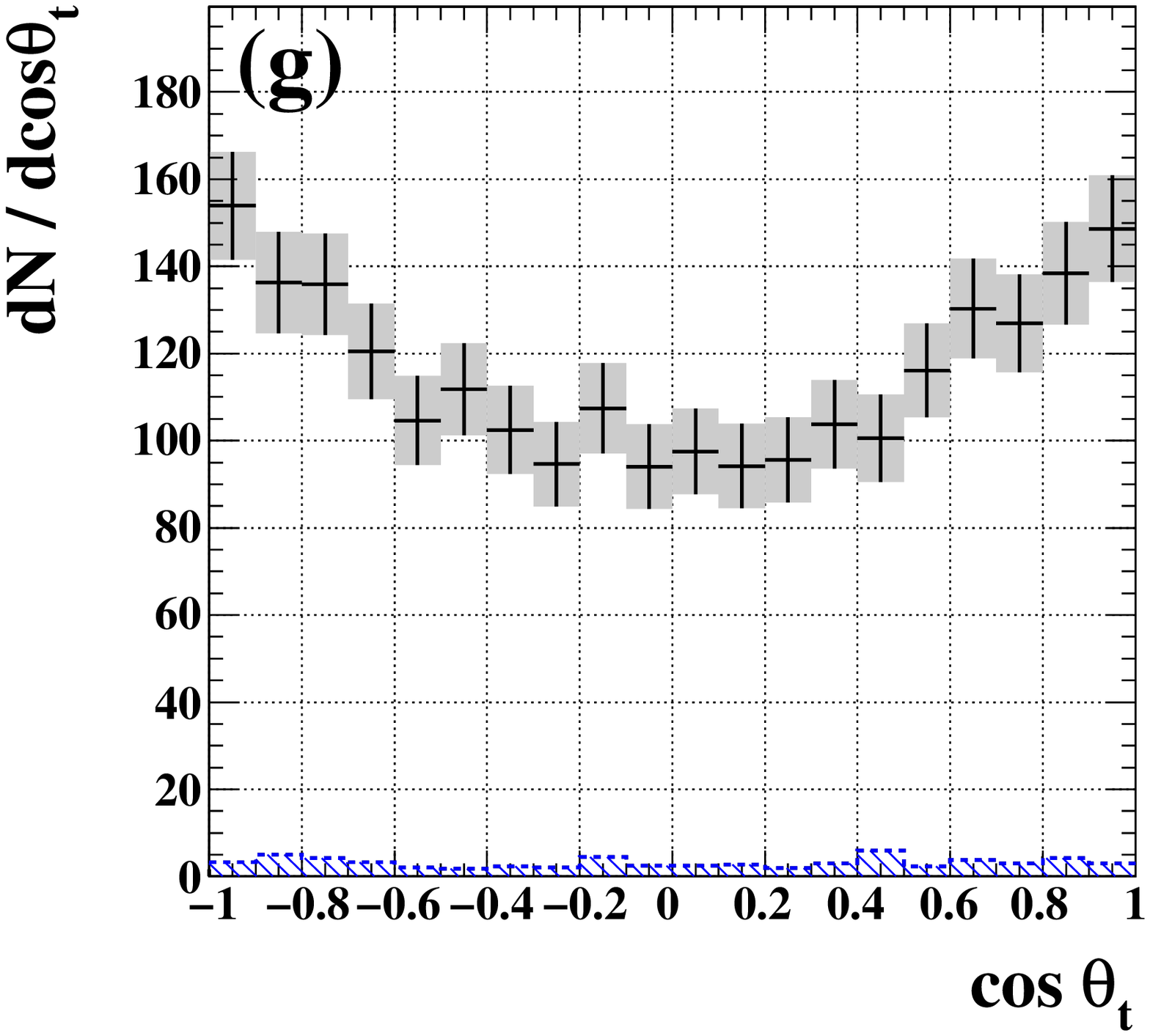}
    \includegraphics[height=4.3cm,clip]{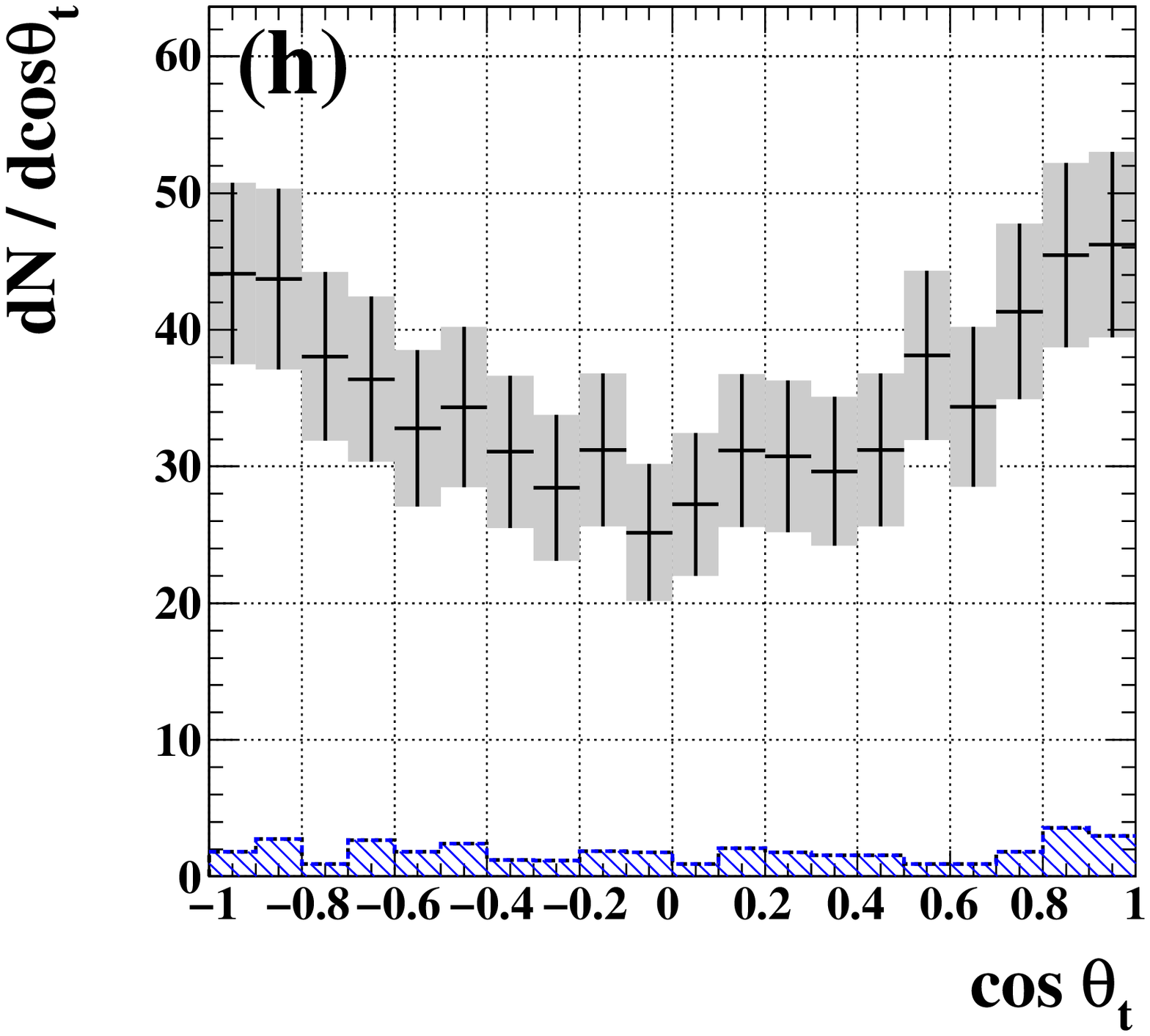}
    \includegraphics[height=4.3cm,clip]{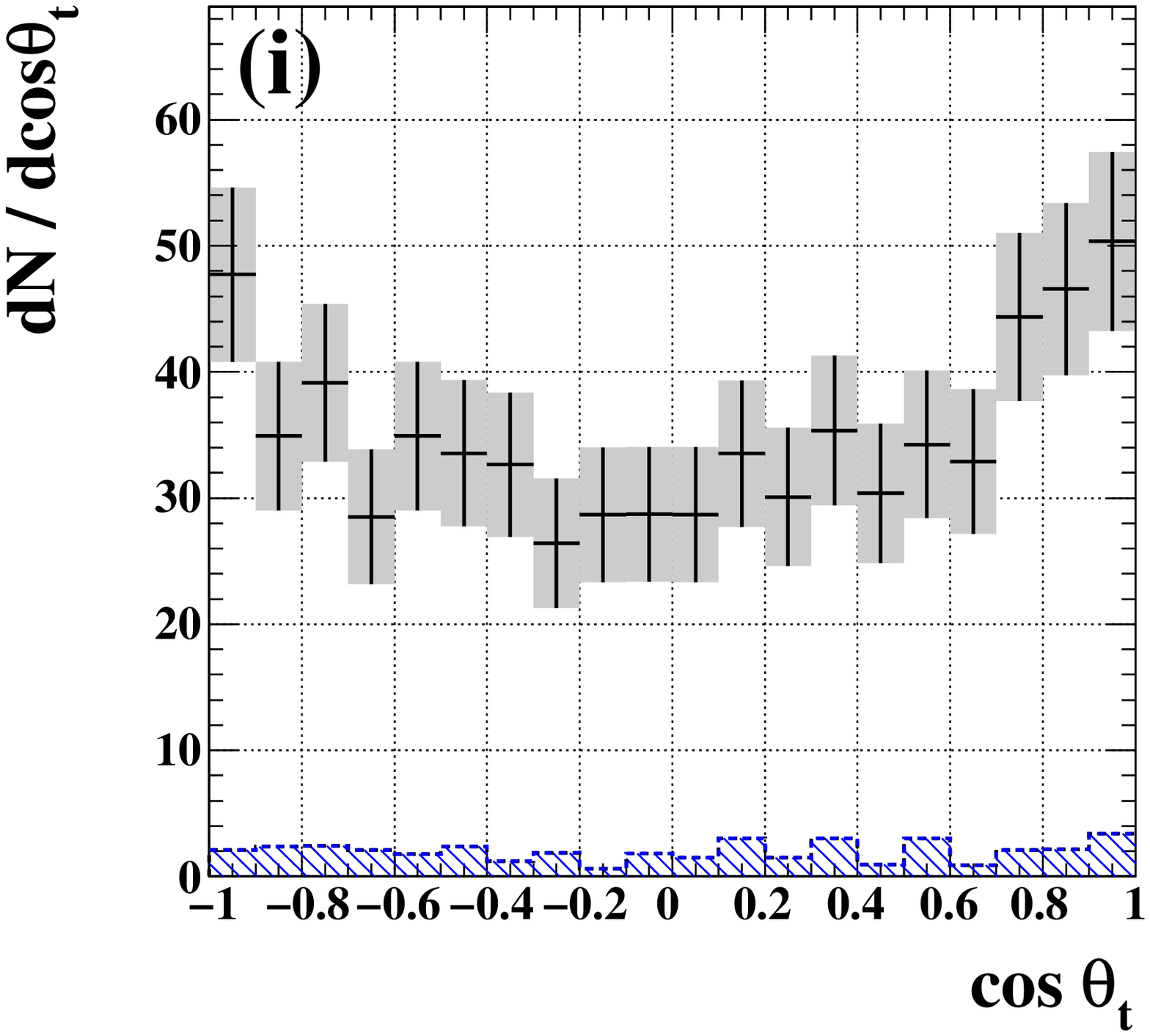}
  \end{center}
  \caption{
        Distributions of the reconstructed polar angle of the identified
        top quark in fully-hadronic $t\bar{t}$ events,
        which are tagged as $(Q_{b_1}, Q_{b_2})$ =
        (a) $(+, -)$, (b) $(+, 0)$, (c) $(0, -)$,
        (d) $(-, +)$, (e) $(-, 0)$, (f) $(0, +)$,
        (g) $(0, 0)$, (h) $(+, +)$, and (i) $(-, -)$
        for the sample in which both $W$ bosons decayed into
        light-quark pairs ($\mathbold{\it bbuddu}$ sample).}
  \label{CosTop1_NoCharm}
\end{figure}

Figures~\ref{CosTop1_NoCharm}-(a) through -(i) show
the distributions of the production angle of $t_1$ for
9 possible combinations of the signs of the vertex charges:
$c_1$ and $c_2$, for the samples where
both $W$ bosons decayed into light quarks.

We can see a clear forward-backward asymmetry
in the case of $C \neq 0$
(Figs.~\ref{CosTop1_NoCharm}-(a) to -(f)).
Notice also that in the first row (-(a) to -(c))
$t_1$ is $t$ while in the second row (-(d) to -(f))
$t_1$ is $\bar{t}$, and hence showing opposite asymmetries.
On the other hand, in the case of $C = 0$, for which
the charge signs of both $b$-jets were undetermined or
they were inconsistent (both $b_1$ and $b_2$ had the same sign),
there is no forward-backward asymmetry visible.
In each figure the contribution from the events with
wrong charge sign is hatched red,
showing an opposite forward-backward asymmetry.
Hatched blue is that from the combinatorial background in which
the reconstructed $b$-jet candidates were not actually $b$-jets,
showing no forward-backward asymmetry.

\begin{figure}[hptb]
  \begin{center}
    \includegraphics[height=4.3cm,clip]{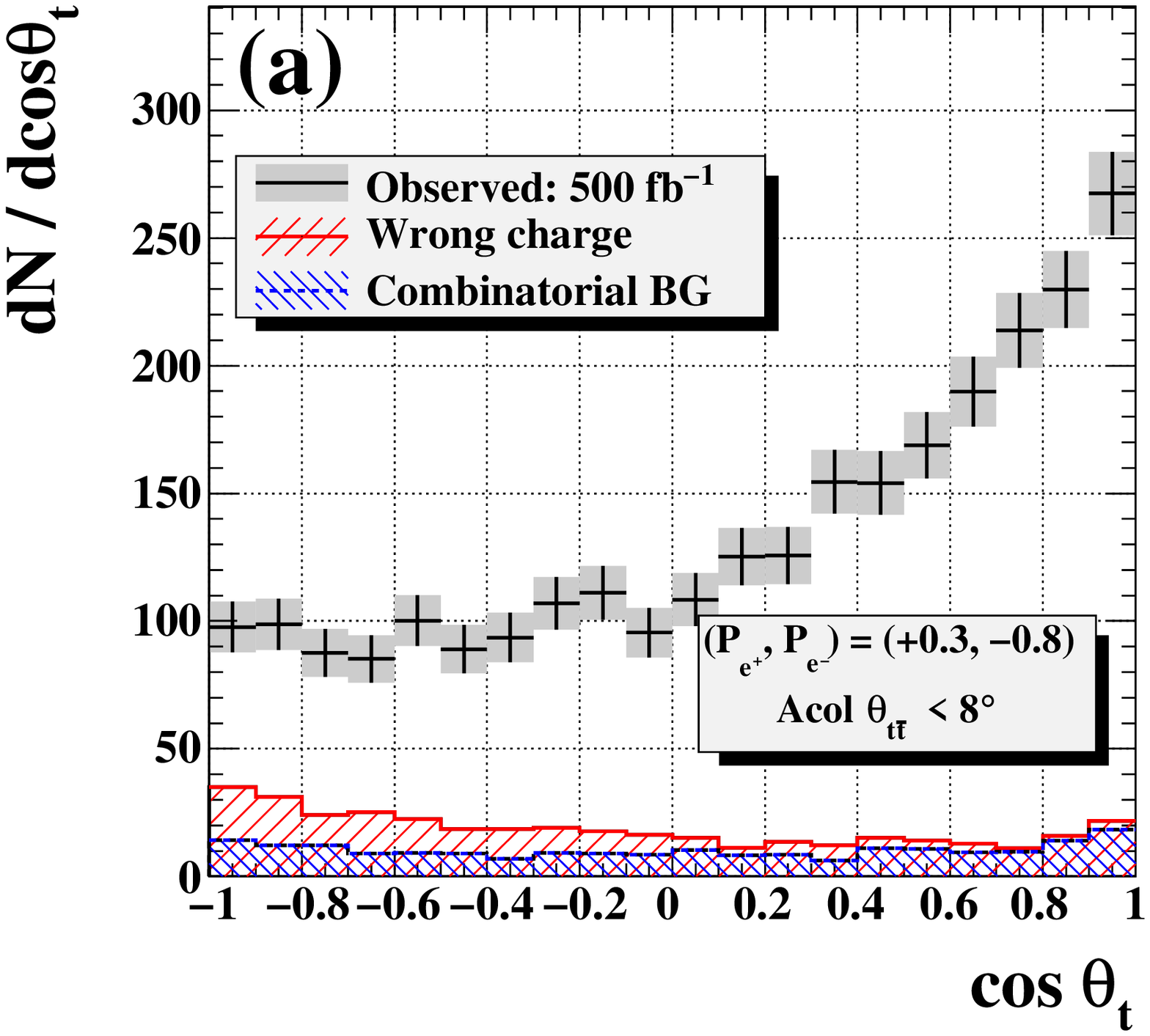}
    \includegraphics[height=4.3cm,clip]{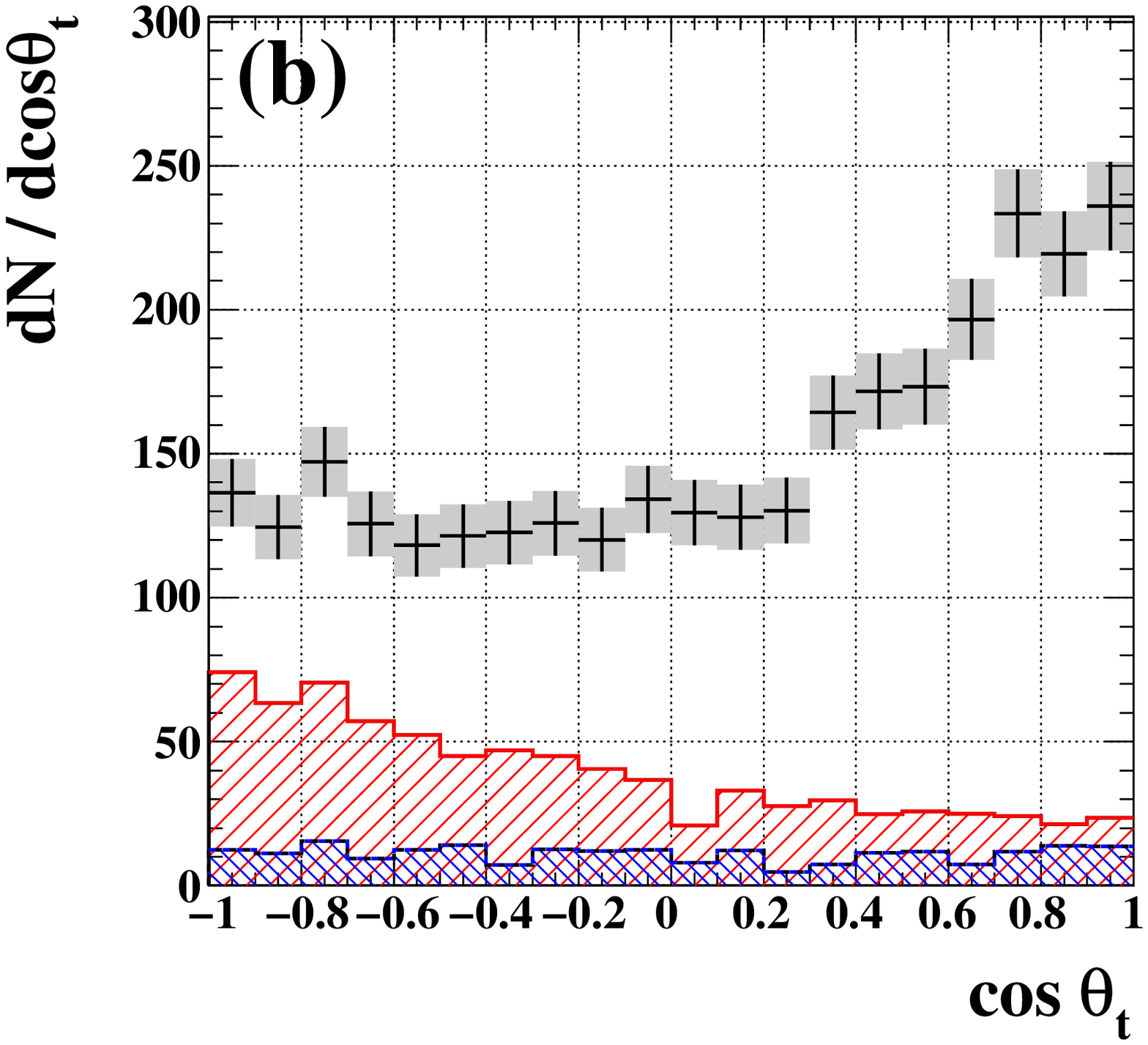}
    \includegraphics[height=4.3cm,clip]{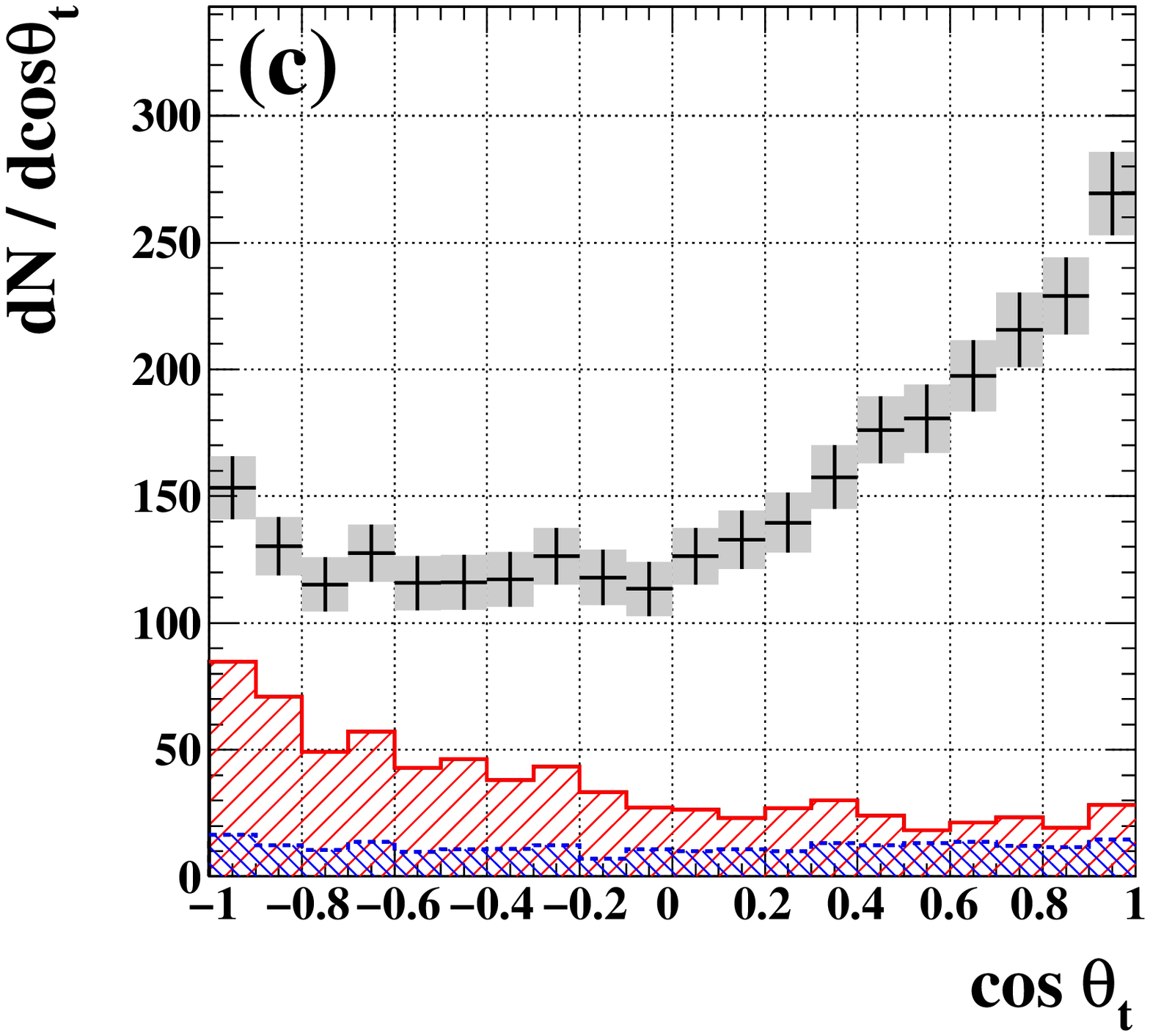}
    \\
    \includegraphics[height=4.3cm,clip]{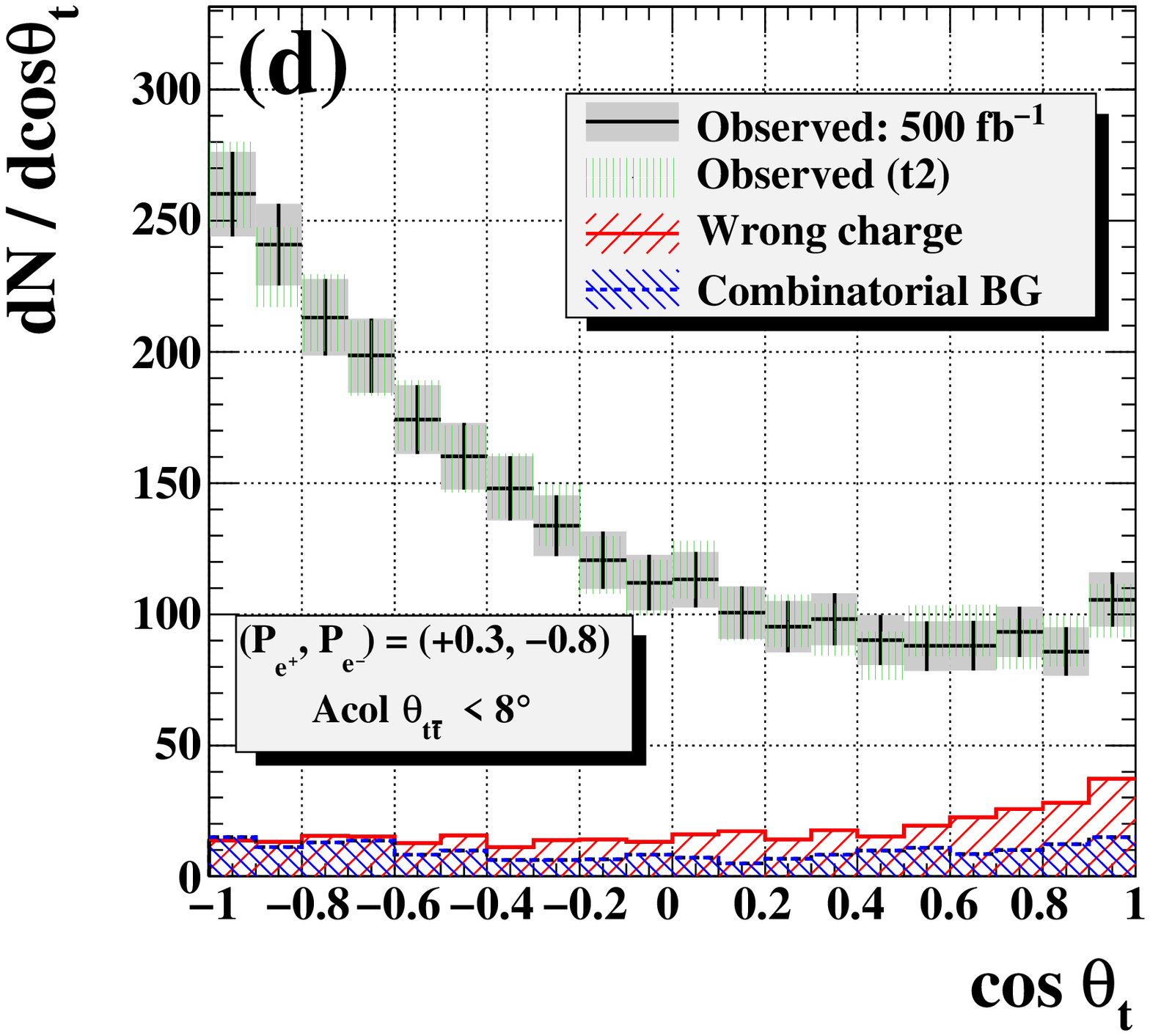}
    \includegraphics[height=4.3cm,clip]{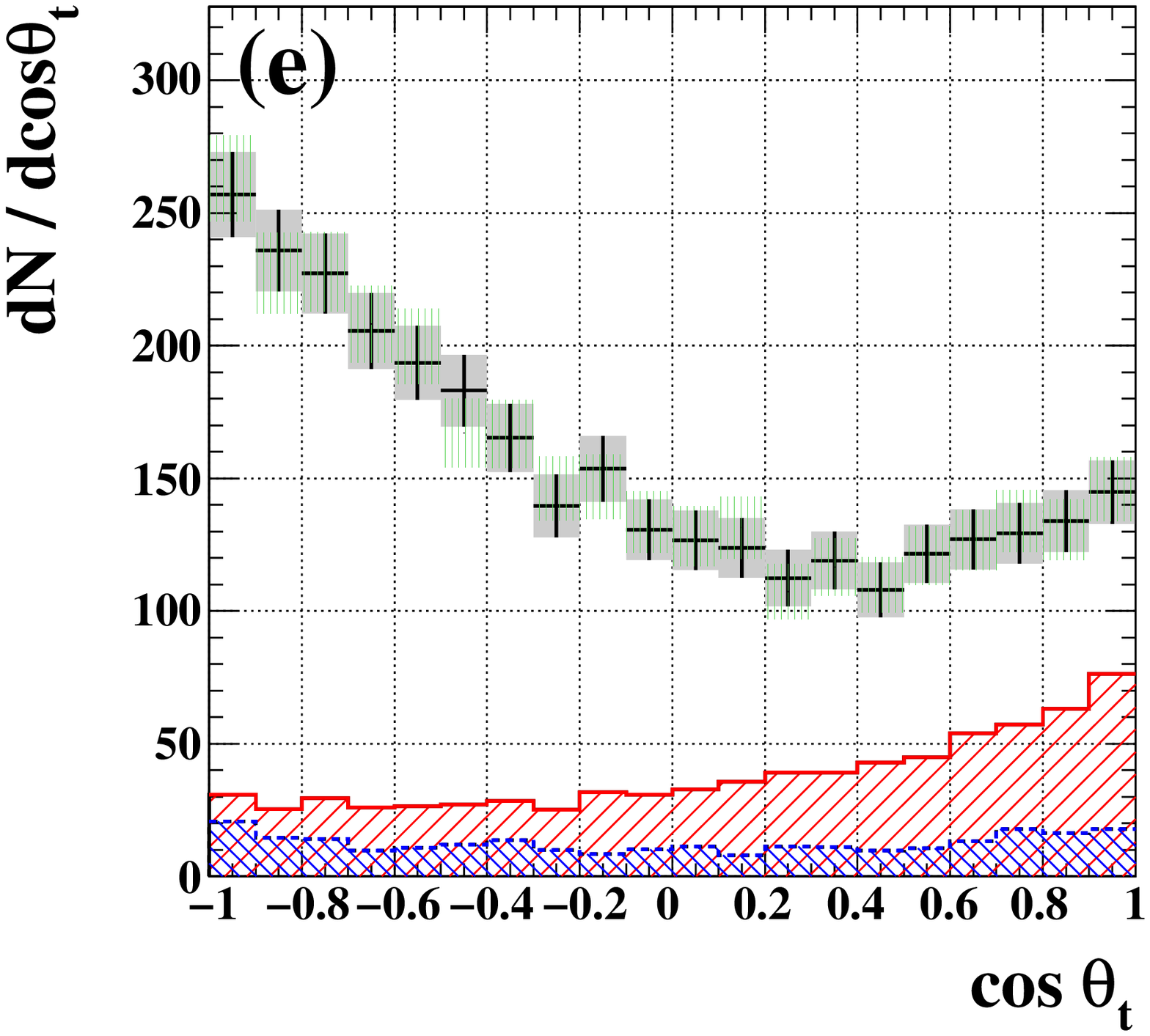}
    \includegraphics[height=4.3cm,clip]{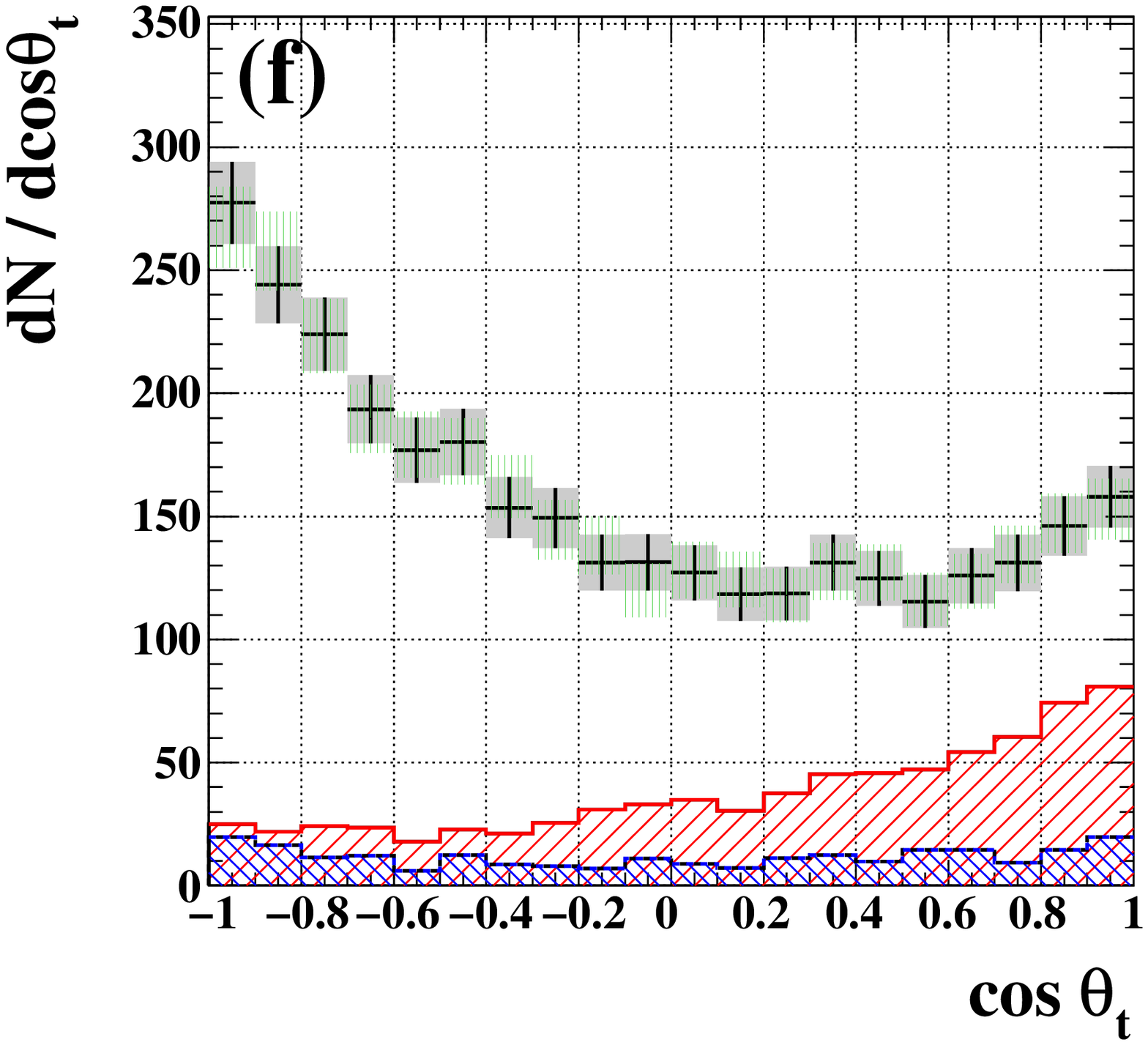}
    \\
    \includegraphics[height=4.3cm,clip]{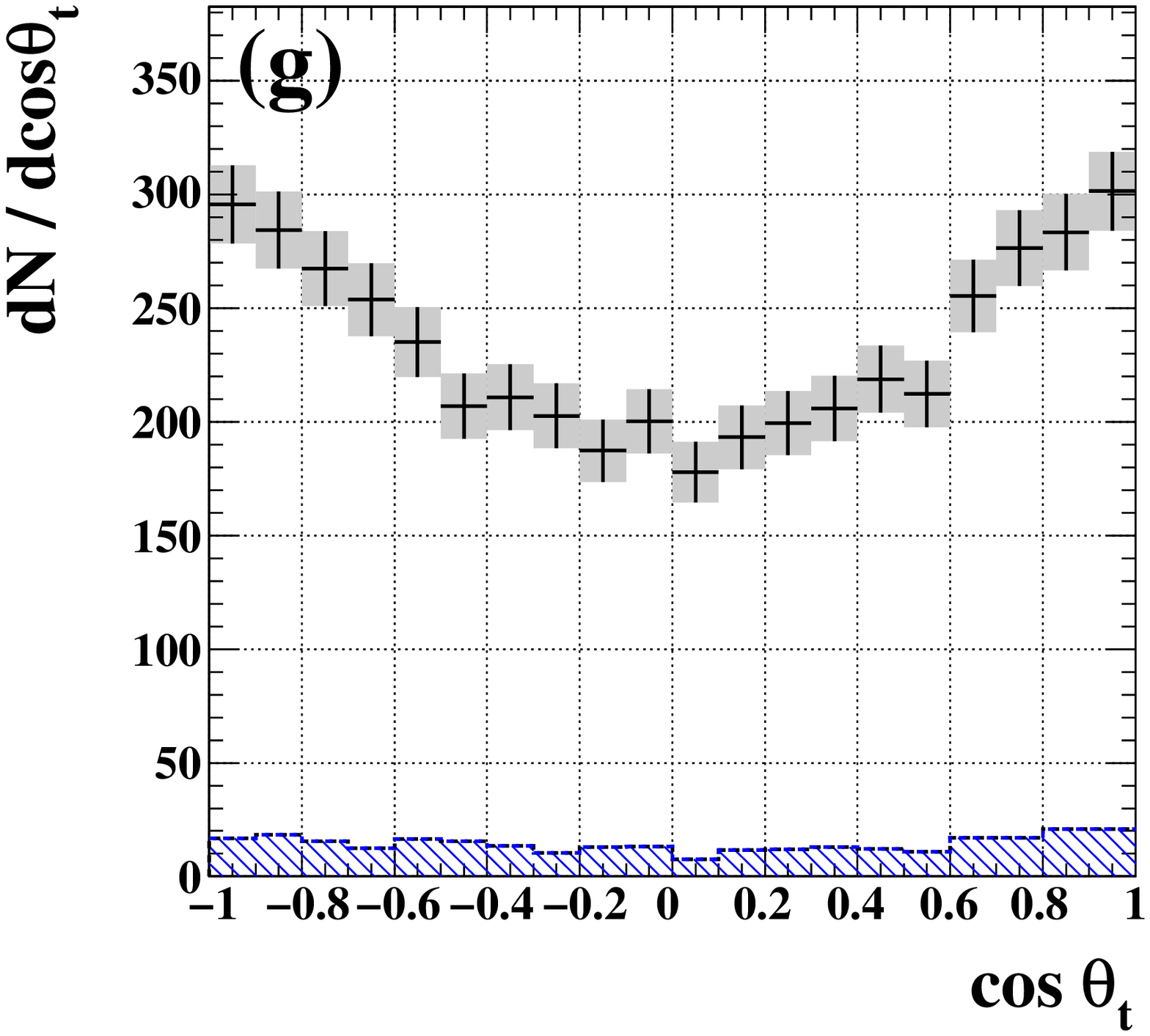}
    \includegraphics[height=4.3cm,clip]{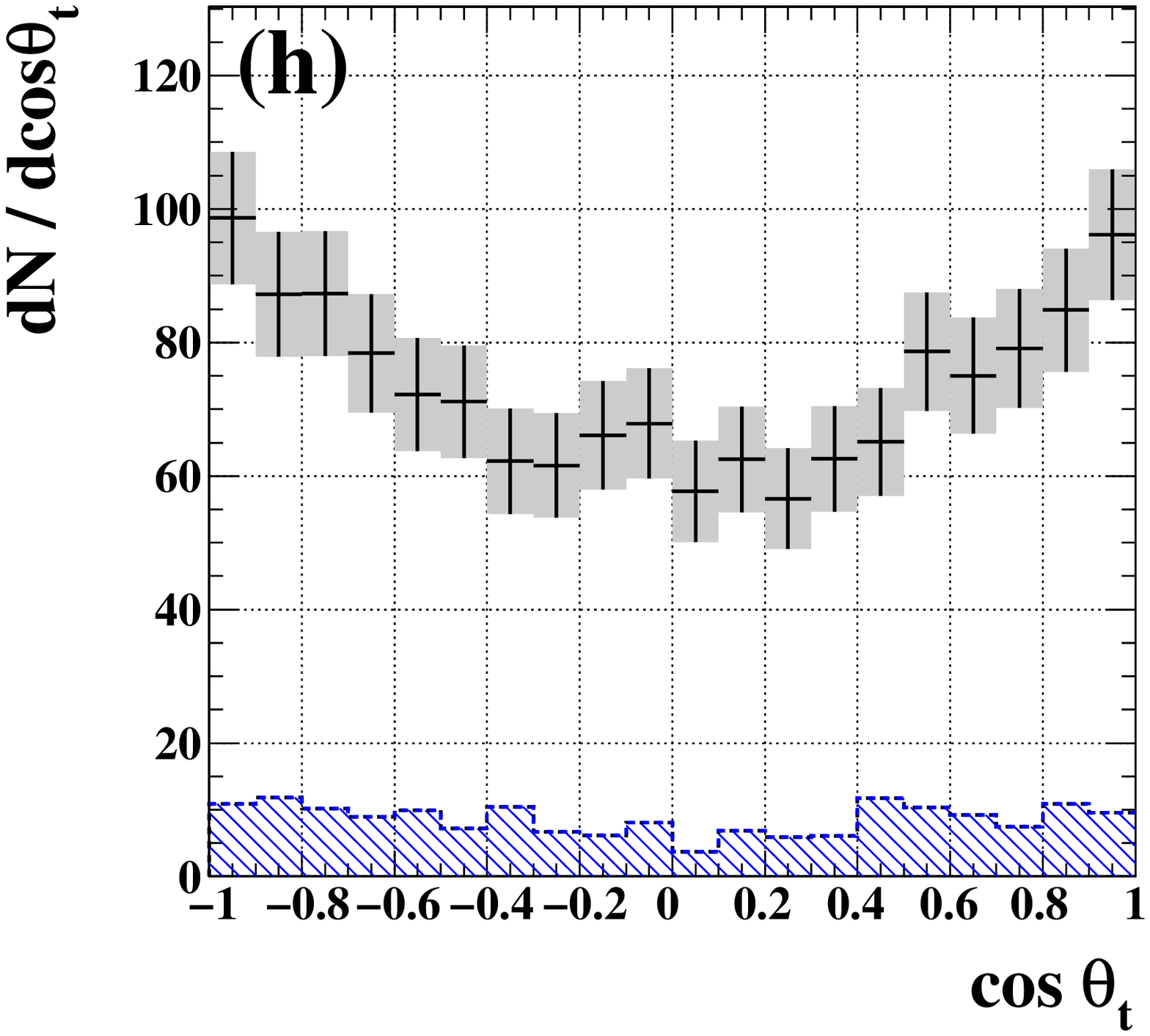}
    \includegraphics[height=4.3cm,clip]{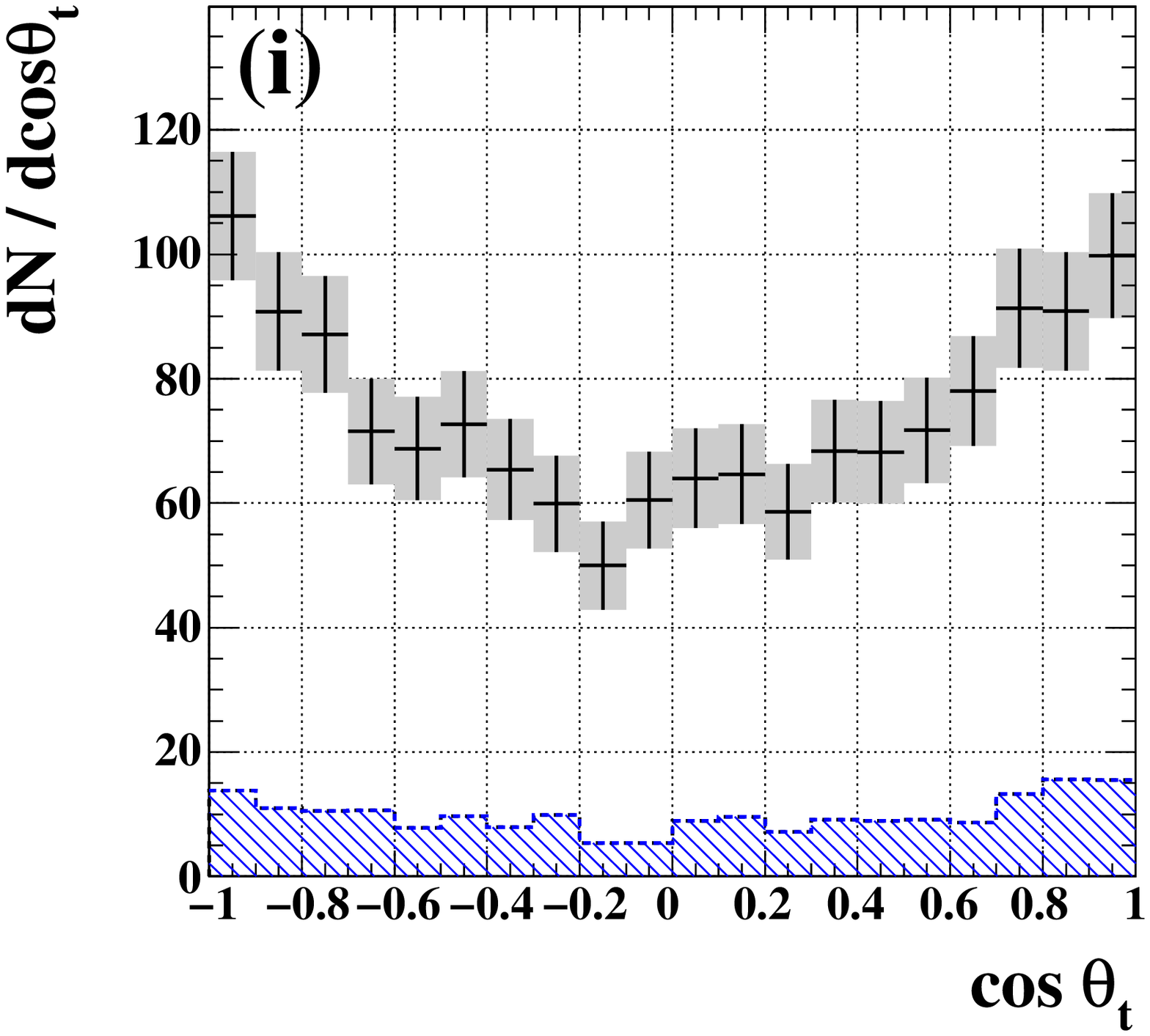}
  \end{center}
  \caption{
        Distributions of the reconstructed polar angle of the identified
        top quark in fully-hadronic $t\bar{t}$ events,
        which are tagged as $(Q_{b_1}, Q_{b_2})$ =
        (a) $(+, -)$, (b) $(+, 0)$, (c) $(0, -)$,
        (d) $(-, +)$, (e) $(-, 0)$, (f) $(0, +)$,
        (g) $(0, 0)$, (h) $(+, +)$, and (i) $(-, -)$
        for the samples in which one of the two $W$ bosons decayed into
        a $c/\bar{c}$-quark ($\mathbold{\it bbudsc/bbcsdu}$ samples).}
  \label{CosTop1_1Charm}
\end{figure}

\begin{figure}[hptb]
  \begin{center}
    \includegraphics[height=4.3cm,clip]{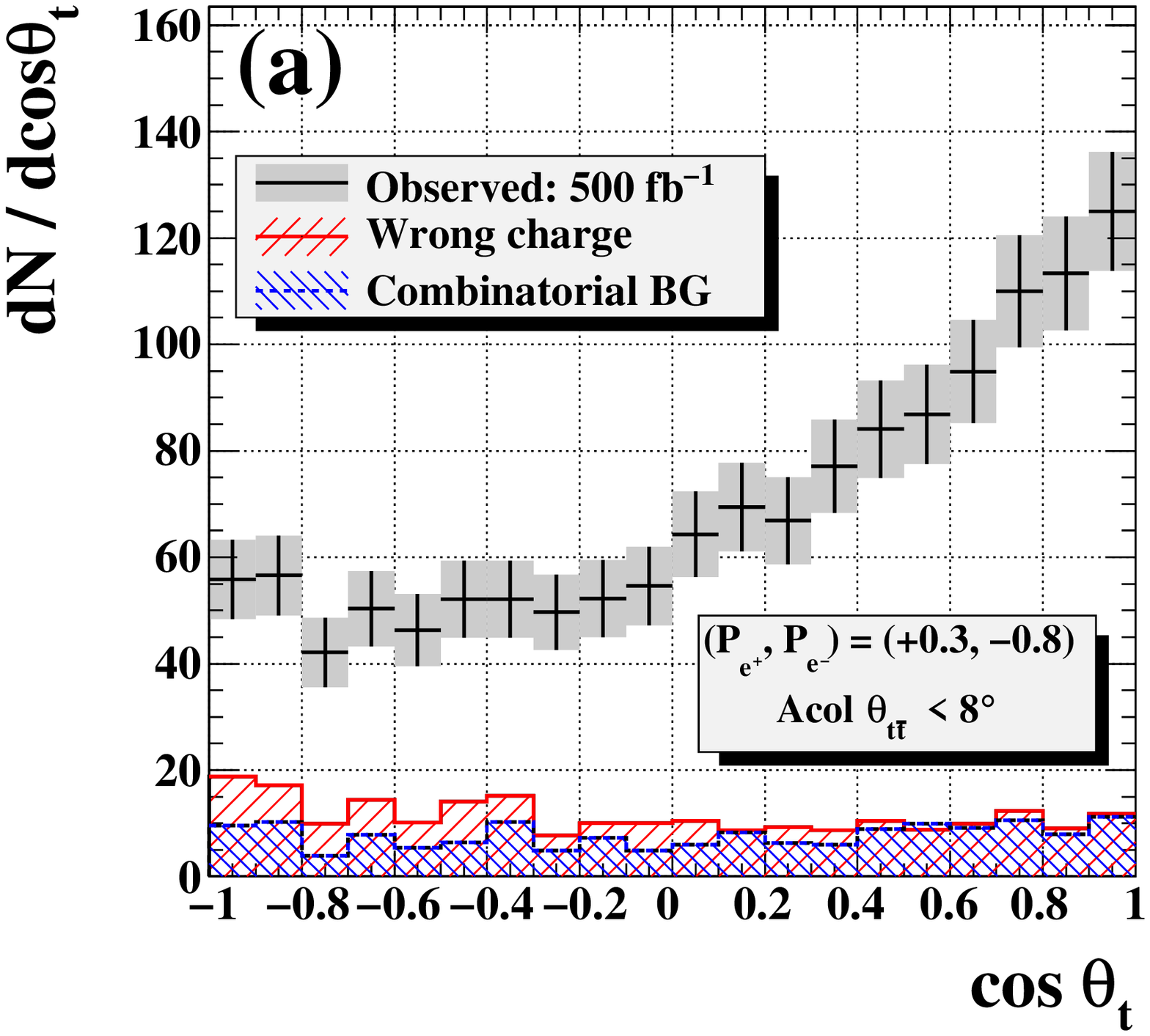}
    \includegraphics[height=4.3cm,clip]{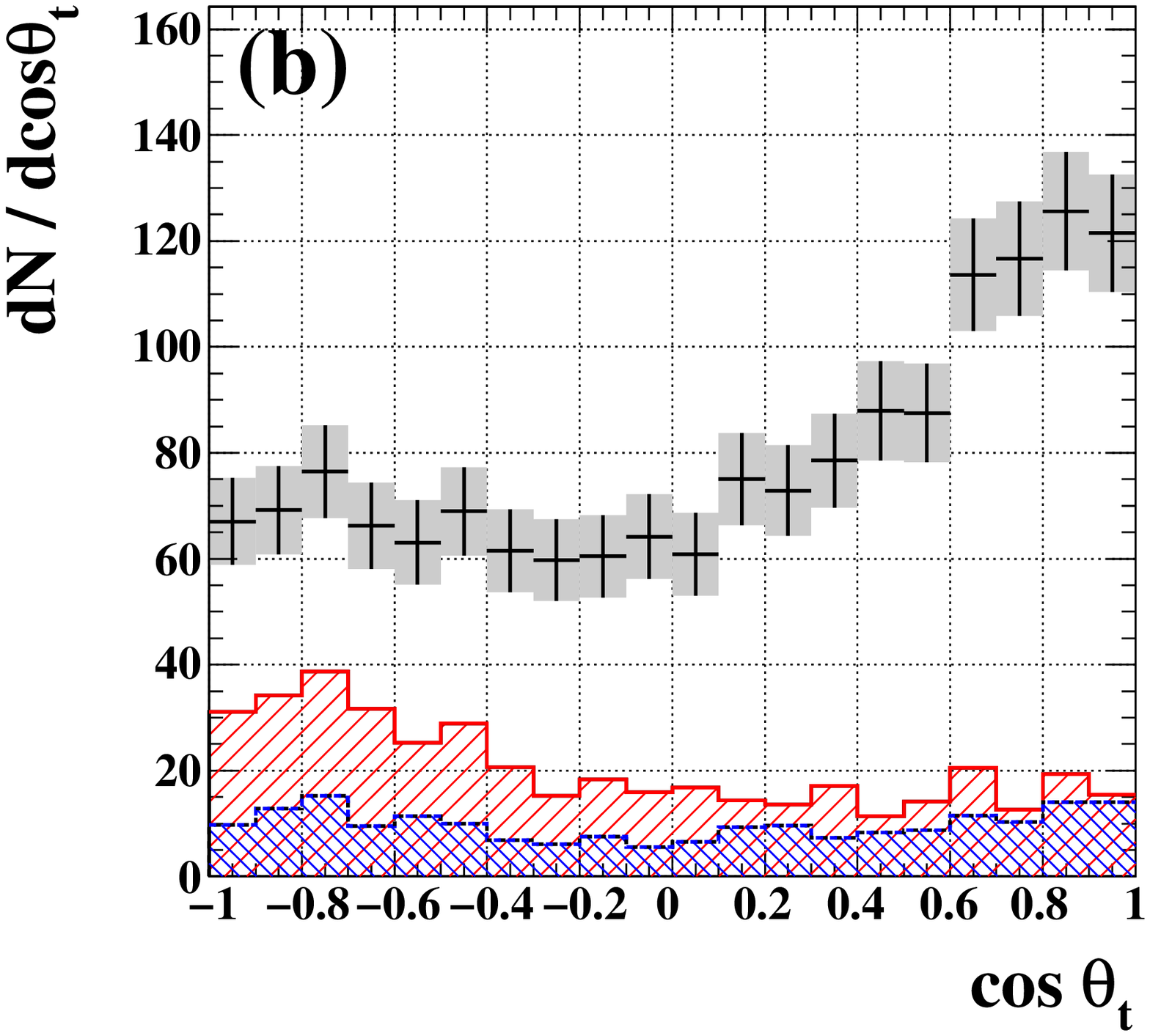}
    \includegraphics[height=4.3cm,clip]{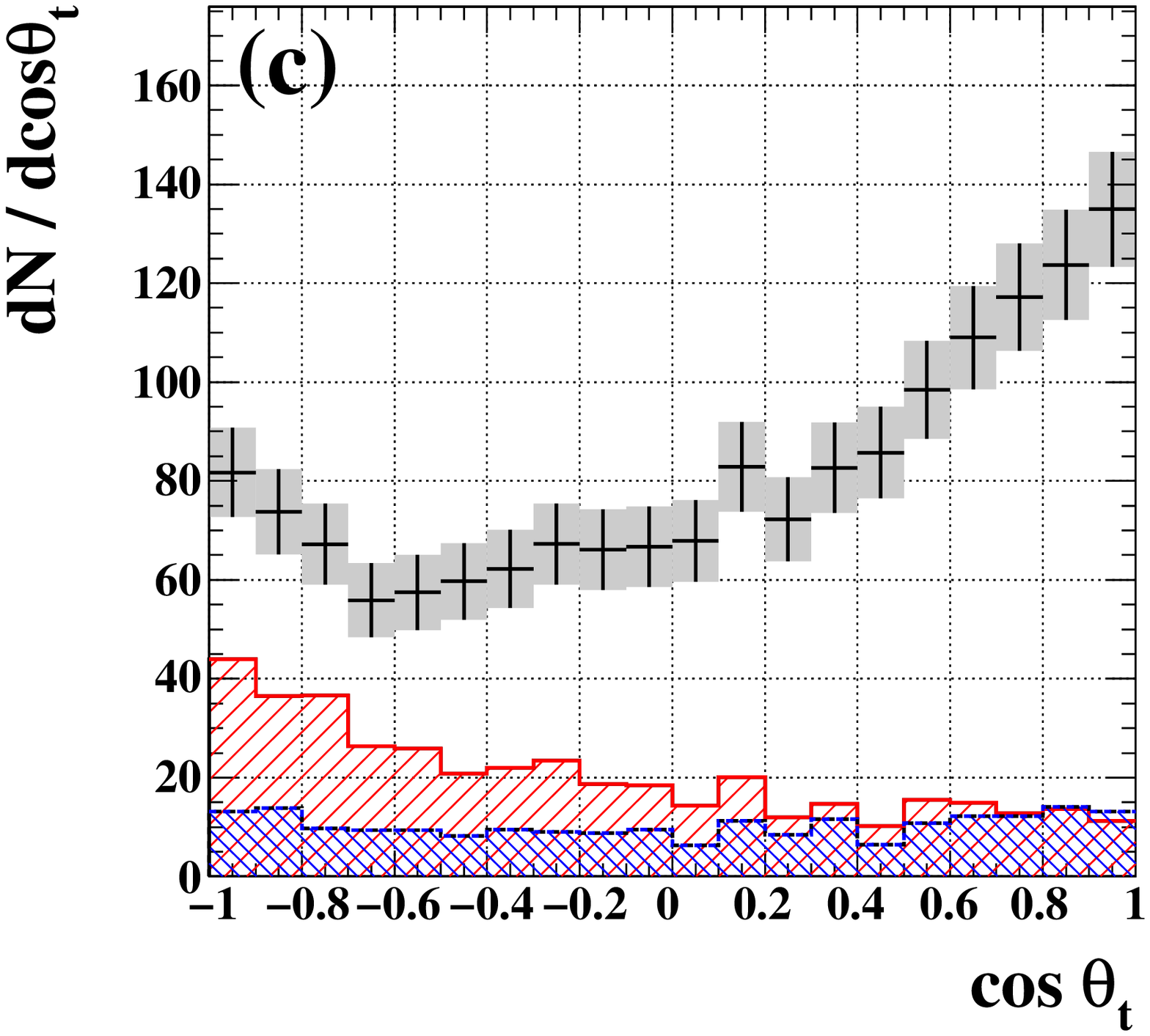}
    \\
    \includegraphics[height=4.3cm,clip]{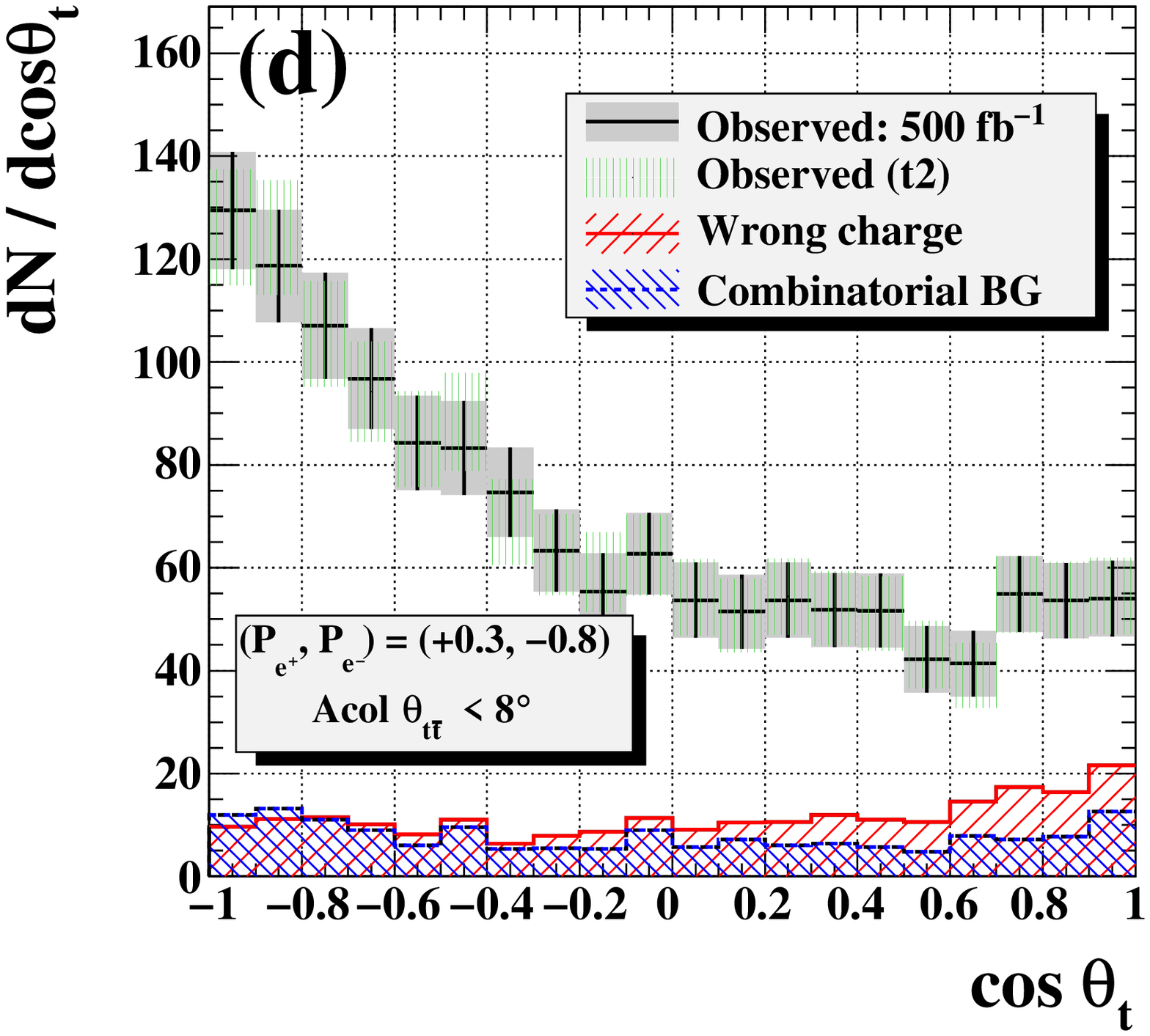}
    \includegraphics[height=4.3cm,clip]{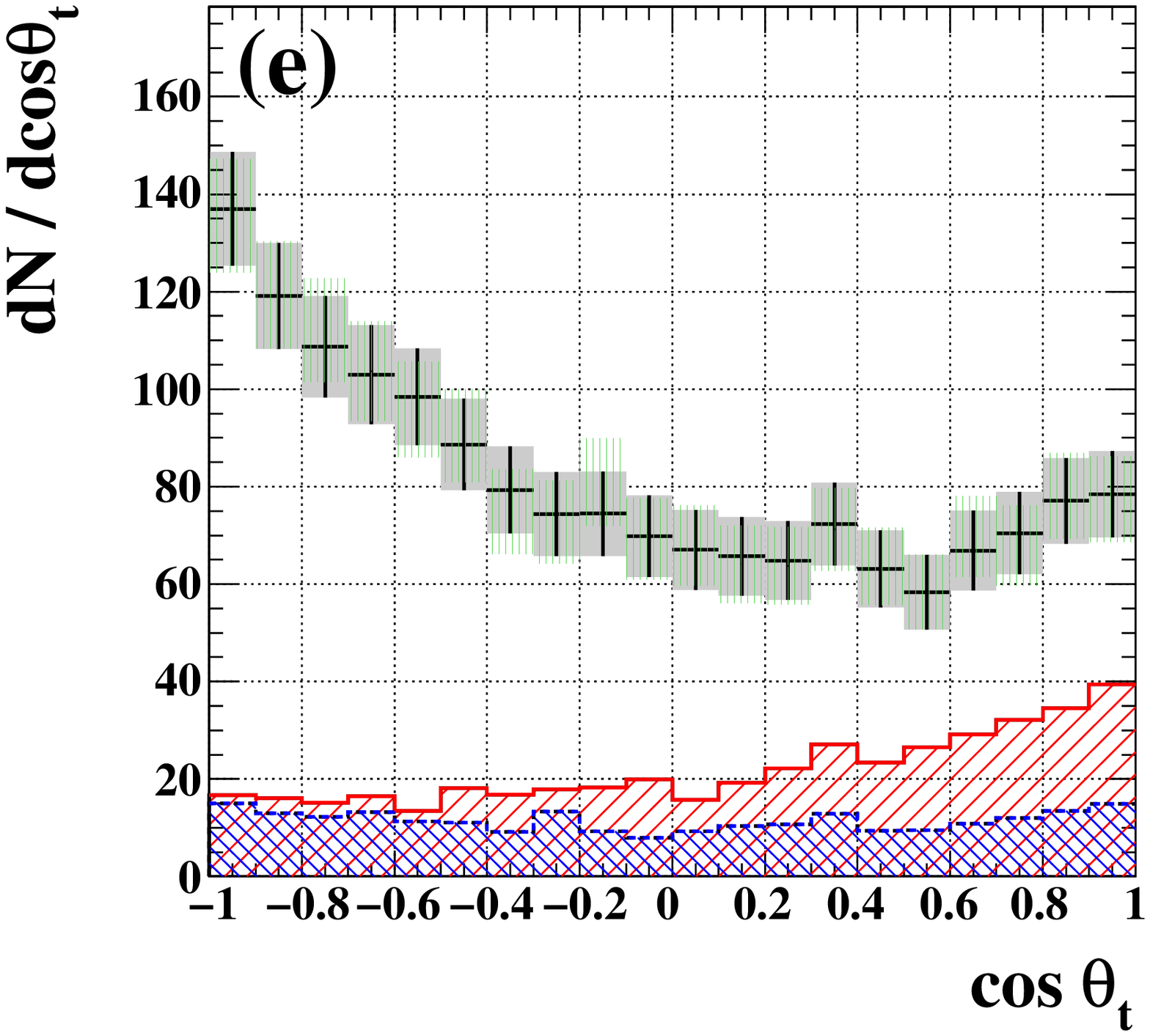}
    \includegraphics[height=4.3cm,clip]{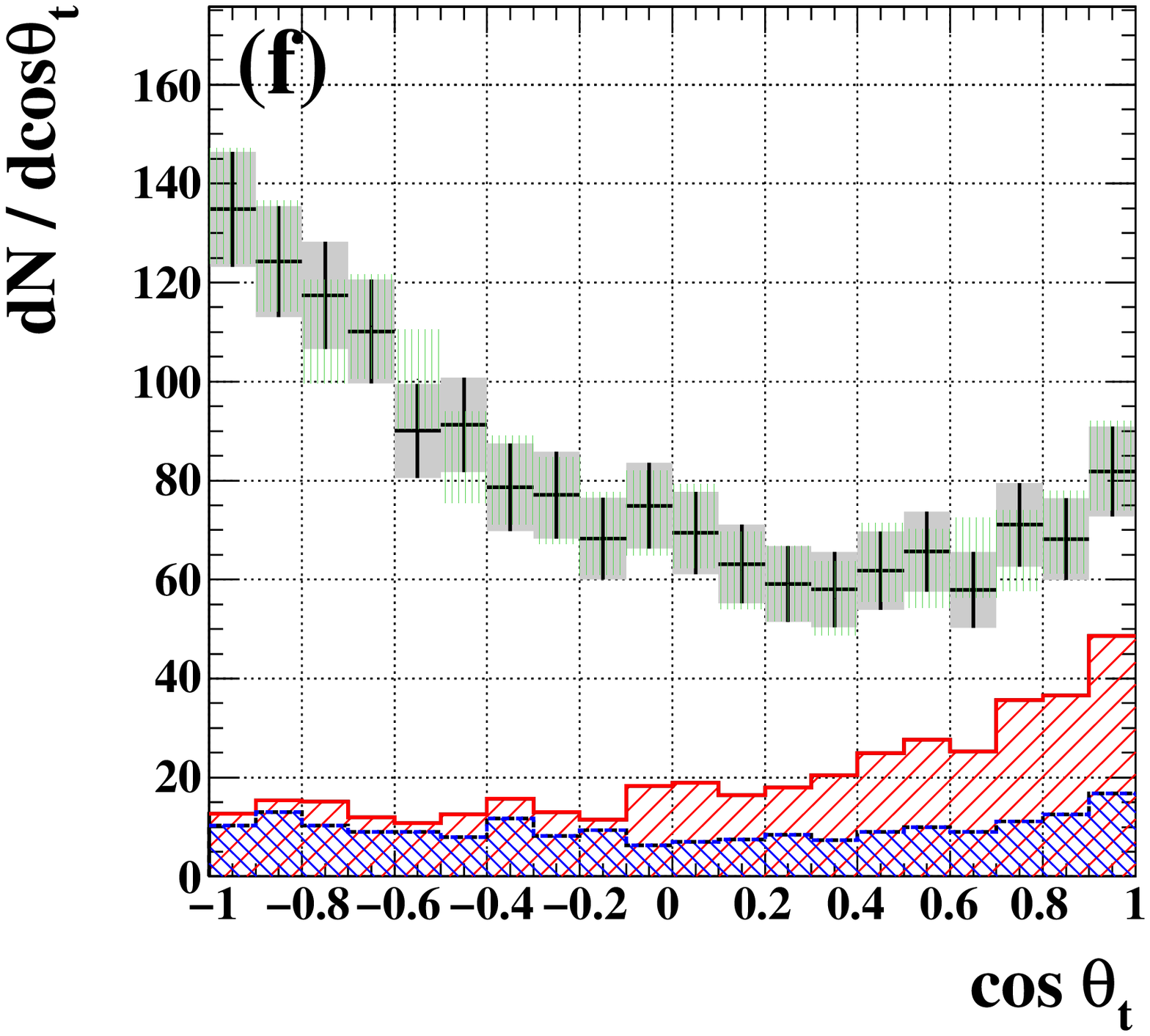}
    \\
    \includegraphics[height=4.3cm,clip]{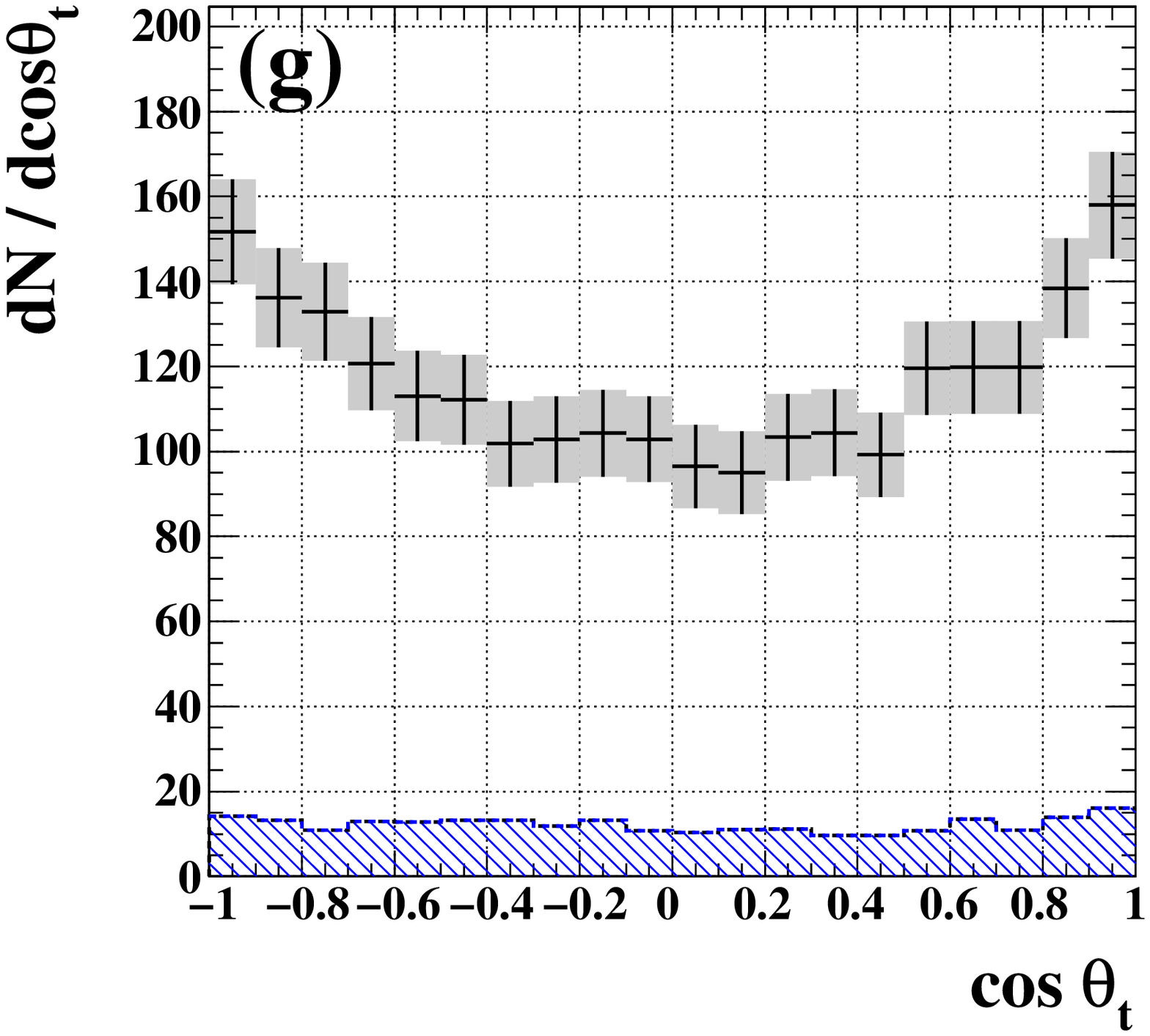}
    \includegraphics[height=4.3cm,clip]{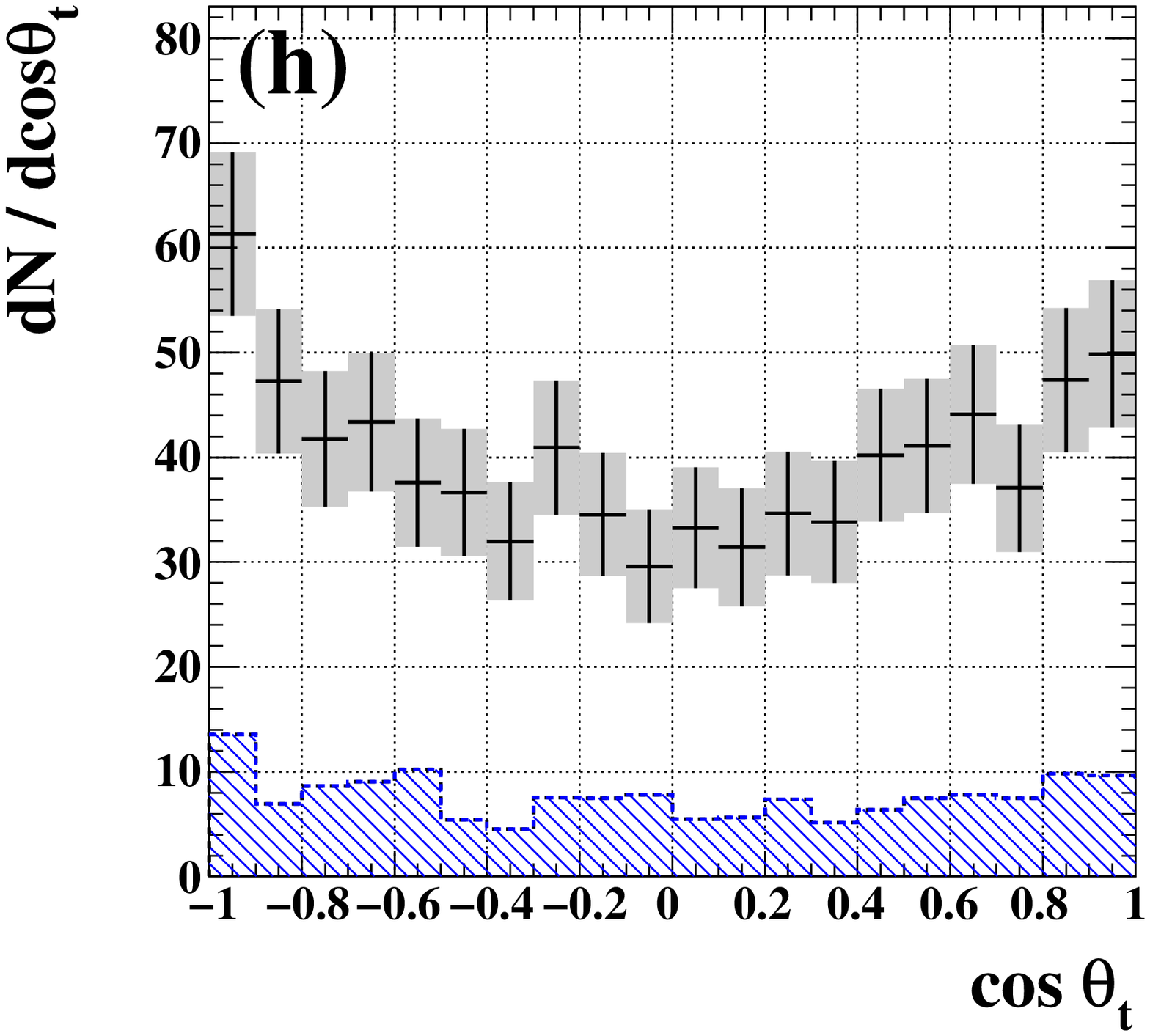}
    \includegraphics[height=4.3cm,clip]{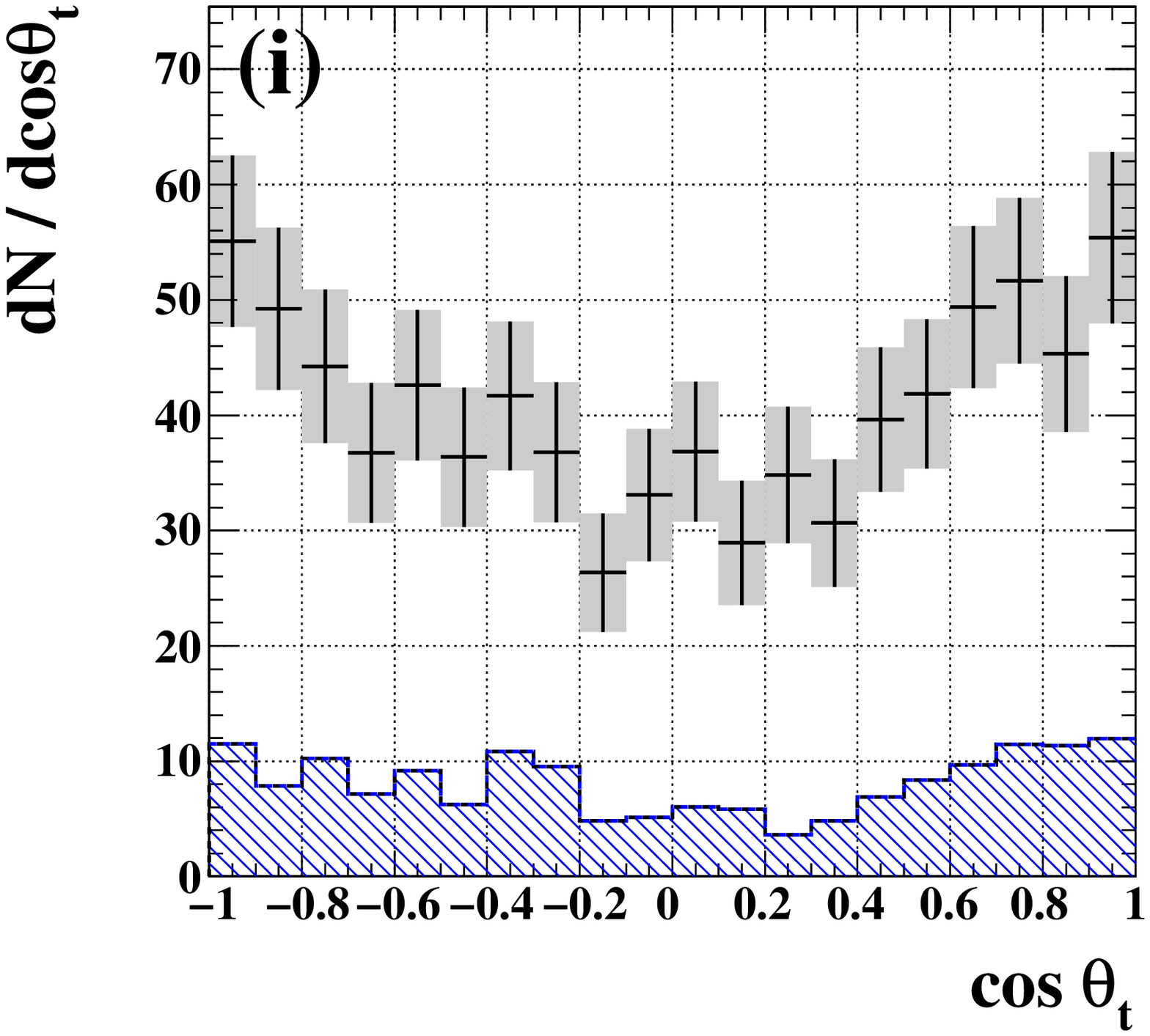}
  \end{center}
  \caption{
        Distributions of the reconstructed polar angle of the identified
        top quark in fully-hadronic $t\bar{t}$ events,
        which are tagged as $(Q_{b_1}, Q_{b_2})$ =
        (a) $(+, -)$, (b) $(+, 0)$, (c) $(0, -)$,
        (d) $(-, +)$, (e) $(-, 0)$, (f) $(0, +)$,
        (g) $(0, 0)$, (h) $(+, +)$, and (i) $(-, -)$
        for the sample in which both $W$ bosons decayed into
        $c/\bar{c}$-quarks ($\mathbold{\it bbcssc}$ sample).}
  \label{CosTop1_2Charm}
\end{figure}

The combinatorial background depends on
the flavour into which $W$ bosons decay.
Figures~\ref{CosTop1_1Charm}-(a) through -(i) are the same figures
as Figs.~\ref{CosTop1_NoCharm}-(a) through -(i),
but plotted for the samples where one of the two $W$ bosons decayed
into a $c/\bar{c}$-quark.
Figures~\ref{CosTop1_2Charm}-(a) through -(i) are similar figures
for the samples where both $W$ bosons decayed
into $c/\bar{c}$-quarks.
We can see clearly that the combinatorial background
grows with the number of $c$-jets in the final states,
since the probability of mis-identifying charm as bottom increases.

If $C$ is positive, $t_1$ is $t$,
while if $C$ is negative, $t_1$ is $\bar{t}$.
If we can assume that $t_1$ and $t_2$ are back-to-back,
the production angle of $t$ is obtained
from the angle of $t_1$ as
\begin{equation}
  cos\theta_t \equiv \sigma_C \cdot |\cos\theta_{t_1}|,
\label{Eq:Cost}
\end{equation}
where $\sigma_C$ is the sign of $C$.
In order to test this assumption
we compared the distributions for $t_1$ with those of $t_2$
after reflection (hatched green) in
Figs.~\ref{CosTop1_NoCharm},~\ref{CosTop1_1Charm},
and~\ref{CosTop1_2Charm}-(d) to -(f).
This comparison confirmed the assumption\footnote{
Strictly speaking, Eq.\,(\ref{Eq:Cost}) does not hold
on an event-by-event basis
because of initial state radiation and beamstrahlung.
On average, however, $t_1$ and $t_2$ are back-to-back,
allowing us to merge the six cases with $C \neq 0$.},
allowing us to use
Eq.\,(\ref{Eq:Cost}) to combine all the figures
with $C \neq 0$.
The selection efficiency of this cut is 69\%.

The differential cross section for $e^+e^- \to t\bar{t}$ and
consequently its forward-backward asymmetry ($A^t_{FB}$) depend on
the center-of-mass energy of the $t\bar{t}$ system,
which, in turn, depends on the amount of initial state radiation
and beamstrahlung.
In order to make $A^t_{FB}$ well-defined,
we hence rejected events with $\sqrt{\hat{s}}$
significantly less than 500 GeV,
by requiring the acollinearity between $t_1$ and $t_2$
to be less than $8^\circ$.
This final cut discarded 24\% of the events so far survived.
The overall selection efficiency is 20\% for fully hadronic
$t\bar{t}$ events.

\begin{figure}[hbtp]
  \begin{center}
    \includegraphics[height=7cm,clip]{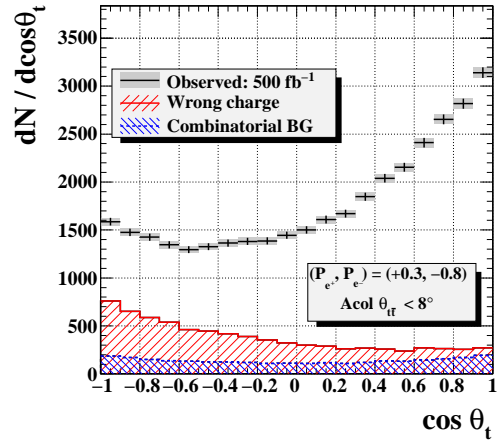}
  \end{center}
  \caption{
        Distribution of the reconstructed polar angle of the identified
        top quark in fully-hadronic $t\bar{t}$ events.
        The contributions from events with the wrong charge (red) and
        the case where the $b$-quark is mis-identified are also shown (blue).}
  \label{CosTop1}
\end{figure}

The resultant production angle distribution is shown
in Fig.~\ref{CosTop1}.
Of the final sample 83\% have
correctly identified signs of top quark charge.

%
%
\section{Determination of $A^t_{FB}$}
\label{Sec:TopAFB}

The production angle distribution Fig.~\ref{CosTop1}
is distorted by the charge mis-identification.
The distortion can be corrected by using
the following formulae:
\begin{equation}
\left\{ {\begin{array}{*{20}c}
   dN_{\rm obs}(\theta) & = &
   p(\theta) \cdot
   \eta(\theta) \cdot
   {\cal L} \cdot d\sigma(\theta)
   +
   \bar{p}(\pi - \theta) \cdot
   \eta(\pi - \theta) \cdot
   {\cal L} \cdot d\sigma(\pi - \theta)
   +
   dB(\theta) \\
   dN_{\rm obs}(\pi - \theta) & = &
   p(\pi - \theta) \cdot
   \eta(\pi - \theta) \cdot
   {\cal L} \cdot d\sigma(\pi - \theta)
   +
   \bar{p}(\theta) \cdot
   \eta(\theta) \cdot
   {\cal L} \cdot d\sigma(\theta)
   +
   dB(\pi - \theta)
\end{array}} \right.
\end{equation}
where $\eta(\theta)$ is the acceptance,
$p(\theta)$ ($\bar{p}(\theta)$) is the probability of
correctly (wrongly) assigning the charge sign
at production angle $\theta$,
and ${\cal L}$ is the integrated luminosity.

\begin{figure}[hbtp]
  \begin{center}
    \includegraphics[height=7cm,clip]{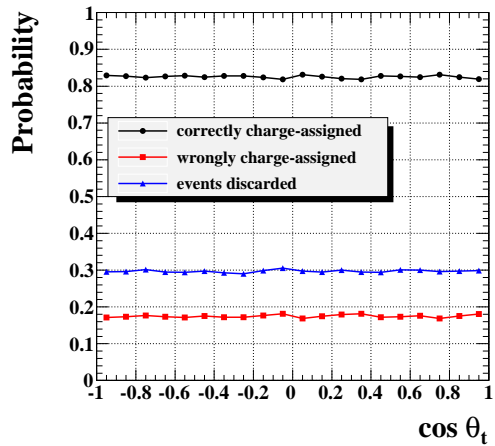}
  \end{center}
  \caption{
        $\cos\theta_{t1}$ dependence of
        the charge-sign identification (black)
        and mis-identification (red) probabilities.}
  \label{ProbChargeID}
\end{figure}

$p(\theta)$ and $\bar{p}(\theta)$ are plotted
in Fig.~\ref{ProbChargeID} as the black and red lines,
respectively.

The figure shows no $\theta$-dependency, allowing us
to set $p(\theta) = p$ and $\bar{p}(\theta) = \bar{p}$.
Solving for $d\sigma(\theta)$, we thus obtain
the following formula for the differential cross section:
\begin{equation}
   d\sigma(\theta) =
   \frac{
     p \cdot (dN_{\rm obs}(\theta) - dB(\theta))
     -
     \bar{p} \cdot (dN_{\rm obs}(\pi - \theta) - dB(\pi - \theta))
   }
   {
     (p^2 - \bar{p}^2) \eta(\theta) \cdot {\cal L}
   }.
\end{equation}

\begin{figure}[hbtp]
  \begin{center}
    \includegraphics[height=7cm,clip]{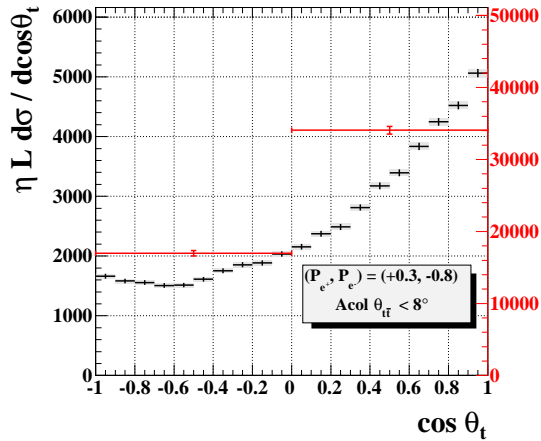}
  \end{center}
  \caption{
        $\cos\theta_{t1}$ distribution after
        the correction for the charge mis-identification.}
  \label{AfbTop}
\end{figure}

Figure~\ref{AfbTop} shows the production angle distribution
after the correction for the charge mis-identification.

The acceptance function $\eta(\theta)$ is in general
$\theta$-dependent.
In our case, however, it turned out at that
the $\theta$-dependence was negligible\footnote{
Notice that for the reconstructed $t$ or $\bar{t}$,
being a 3-jet system, there is no acceptance hole
even at $\theta = 0$ or $\pi$.}.
The acceptance hence cancels out in the calculation of
the forward-backward asymmetry, resulting in
\begin{equation}
   A^{t}_{FB} =
   \frac{
     \int_{0 < \theta < \pi/2} dN(\theta)
     -
     \int_{\pi/2 < \theta < \pi} dN(\theta)
   }
   {
     \int_{0 < \theta < \pi/2} dN(\theta)
     +
     \int_{\pi/2 < \theta < \pi} dN(\theta)
   },
\end{equation}
where $dN(\theta) \equiv \eta \cdot {\cal L}
\cdot d\sigma(\theta)$.

From Fig.~\ref{AfbTop} we finally obtain
\begin{equation}
   A^{t}_{FB} = 0.334 \pm 0.0079 \mbox{ (stat.)},
\end{equation}
where the center of mass energy of $\sqrt{s}=500$ GeV,
an integrated luminosity of $500 \, {\rm fb}^{-1}$,
and a beam polarization combination of
$P(e^+, e^-) = (+30\%, -80\%)$.

%
%
\section{Summary}
\label{Sec:Summary}

We have studied the measurement accuracy
of the forward-backward asymmetry with the ILD detector
for the $e^+e^- \rightarrow t\bar{t}$ process
in the 6-jet mode at $\sqrt{s} = 500$ GeV.
In the analysis the vertex charges of $b$-jets were used
to identify $t$ and $\bar{t}$.
The efficiency to tag the vertex charge of a $b$-jet
was 28\% with the purity of 75\%.
Having two $b$-jets in each event,
the probability to identify $t$ and $\bar{t}$ in the event was
71\% with the probability of wrong charge assignment of 12\%.
The measured angular distribution was corrected
for wrong $t/\bar{t}$ charge assignments.
From the number of events in forward and backward hemispheres
after the correction, we obtained $A^{t}_{FB} = 0.334 \pm 0.0079$,
where quoted error is statistical only.

%
%
\vskip 0.5cm
\begin{flushleft}
\underline{\bf{Acknowledgments}}
\end{flushleft}

The authors wish to thank all the members of the ILD optimization group
for useful discussions and comments.
This study is supported in part by the Creative Scientific Research
Grant No. 18GS0202 of the Japan Society for Promotion of Science (JSPS)
and the JSPS Core University Program.

\begin{footnotesize}
%
%
\def\plb#1#2#3{{\it Phys.~Lett.~}{\bf B#1}, #2 (#3)}
\def\prd#1#2#3{{\it Phys.~Rev.~}{\bf D#1}, #2 (#3)}
\def\prl#1#2#3{{\it Phys.~Rev.~Lett.~}{\bf #1}, #2 (#3)}
\def\zpc#1#2#3{{\it Z.~Phys.~}{\bf C#1}, #2 (#3)}

\end{footnotesize}

\end{document}